\documentclass[prb,aps,twocolumn,groupedaddress,floatfix,citeautoscript,showpacs]{revtex4}
\usepackage{graphicx,rotating,subfigure,amsmath,amsfonts,amssymb,delarray}
\usepackage{dsfont}
\usepackage[T1]{fontenc}
\numberwithin{equation}{section}

\newcommand{\e}{\text{e}}
\newcommand{\im}{\text{i}}

\def\12{\frac{1}{2}}

\newcommand{\be}{\begin{equation}}
\newcommand{\ee}{\end{equation}}
\newcommand{\bea}{\begin{eqnarray}}
\newcommand{\eea}{\end{eqnarray}}

\DeclareMathOperator\Tr{Tr}

\allowdisplaybreaks

\begin{document}
\bibliographystyle{apsrev}
\title{Dynamical quantum phase transitions and the Loschmidt echo:\\ A transfer matrix approach}

\author{F. Andraschko}
\affiliation{Department of Physics and Research Center OPTIMAS,
  Technical University Kaiserslautern,
   D-67663 Kaiserslautern, Germany}
\author{J. Sirker}
\affiliation{Department of Physics and Research Center OPTIMAS,
  Technical University Kaiserslautern,
  D-67663 Kaiserslautern, Germany}

\date{\today}

\begin{abstract}
  A boundary transfer matrix formulation allows to calculate the
  Loschmidt echo for one-dimensional quantum systems in the
  thermodynamic limit. We show that nonanalyticities in the Loschmidt
  echo and zeros for the Loschmidt amplitude in the complex plane
  (Fisher zeros) are caused by a crossing of eigenvalues in the
  spectrum of the transfer matrix. Using a density-matrix
  renormalization group algorithm applied to these transfer matrices
  we numerically investigate the Loschmidt echo and the Fisher zeros
  for quantum quenches in the XXZ model with a uniform and a
  staggered magnetic field. We give examples---both in the integrable
  and the nonintegrable cases---where the Loschmidt echo does not show
  nonanalyticities although the quench leads across an equilibrium
  phase transition, and examples where nonanalyticities appear for
  quenches within the same phase. For a quench to the free fermion
  point, we analytically show that the Fisher zeros sensitively depend
  on the initial state and can lie exactly on the real axis already
  for finite system size.  Furthermore, we use bosonization to analyze
  our numerical results for quenches within the Luttinger liquid
  phase.
 \end{abstract}

\pacs{05.70.Ln,64.70.Tg,75.10.Pq}

\maketitle

\section{Introduction}
\label{Intro}
The time evolution of a quantum system is governed by a unitary
operator $U(t)$. As a consequence, the evolution of a quantum state in
time, $|\Psi(t)\rangle =U(t)|\Psi_0\rangle$, where $|\Psi_0\rangle$ is
the initial state, can be reversed $|\Psi_0\rangle
=U^\dagger(t)|\Psi(t)\rangle$. The question how stable such a time
reversal is with respect to small perturbations is of fundamental
importance to understand decoherence, which leads to the emergence of
classical behavior and sets limits for quantum
computation.\cite{Zurek_decoherence,DiVincenzo} For a many-body system,
one might expect on general grounds an inherent instability related to
the dynamical complexity of the problem and quantum
chaos.\cite{Zurek_chaos,JacquodSilvestrov,JalabertPastawski} 

A measure of the stability of time reversal is the Loschmidt echo
defined by\cite{Peres_Loschmidt}
\begin{equation}
\label{Lo_general}
\mathcal{L}(t)=|\langle\Psi_0|e^{iH(g_1)t}e^{-iH(g_0)t}|\Psi_0\rangle|^2.
\end{equation}
Here, $H(g)$ is a time-independent Hamiltonian which is a function of a
microscopic parameter $g$ whose value is assumed to change from $g_0$
during the forward time evolution to $g_1$ during the backward time
evolution. $\mathcal{L}(t)$ thus is nothing but the fidelity between
the state time-evolved with $H(g_0)$ and the state time-evolved with
$H(g_1)$.\cite{GorinProsen}
An advantage of the Loschmidt echo in comparison to many other quantum
information measures is that it is accessible experimentally, for
example, by nuclear magnetic
resonance.\cite{LevsteinUsaj,PastawskiLevstein}

In this paper, we are interested in the recent idea to use the
Loschmidt echo to study and classify quenches in closed quantum
systems. We concentrate on the case where $|\Psi_0\rangle$ is the
ground state of $H(g_1)$. In this case, the Loschmidt echo reduces to
\begin{equation}
\label{Lo_spec}
L(t)=|\langle\Psi_0|e^{-iHt}|\Psi_0\rangle|^2,
\end{equation}
where we have set $H\equiv H(g_0)$. The Loschmidt echo can be
calculated exactly for bilinear Hamiltonians such as the transverse
Ising model.\cite{QuanSong,HeylPolkovnikov} In
Ref.~\onlinecite{HeylPolkovnikov}, it has been, furthermore, pointed
out that the Loschmidt amplitude $Z(z)$ has the form of a partition
function with boundaries fixed by the initial state,
\begin{equation}
\label{Zz}
Z(z)=\langle\Psi_0|e^{-zH}|\Psi_0\rangle \, ,
\end{equation}
with $z\in\mathbb{C}$. Here, a general complex parameter $z$ allows to
investigate the analytic structure of $Z(z)$ in the whole complex
plane.

The motivation for such an approach are studies of the analytic
properties of the thermodynamic grand canonical partition function
$Z_\textrm{th}(\beta,\mu)=\Tr\,\e^{-\beta(H-\mu N)}$ for complex
fugacity $\tilde{z}=\exp(\beta\mu)$ ($\beta$ is real) or of the canonical
partition function $Z_\textrm{th}(\beta)=\Tr\,\e^{-\beta H}$ for
complex inverse temperature $\beta$, which have helped to understand
and classify phase transitions.\cite{BenaDroz} For $\beta$ and $\tilde
z$ real, the grand canonical partition function is a sum of positive
terms and thus strictly positive. Furthermore, $Z_{th}$ is a sum of
analytic terms and therefore an analytic function for any finite
system size.  Phase transitions, signalled by nonanalyticities in the
corresponding thermodynamic potential, can therefore only occur in the
thermodynamic limit. In this limit, zeros in the complex $\tilde z$ or
complex $\beta$ plane will close in on the real axis at the transition
point so that the thermodynamic potential becomes nonanalytic.

The grand canonical partition function $Z_\textrm{th}(\beta,\mu)$ is a
polynomial in the complex fugacity $\tilde z$ with roots which appear
in complex conjugated pairs. At the transition temperature
$T_c=1/\beta_c$ these roots close in on the real axis and the density of
roots determines the order of the phase transition. For the Ising
model in a magnetic field $B$, Lee and Yang showed that the {\it
  Lee-Yang zeros} $\tilde z_j$ of the partition function for complex
fugacity $\tilde z=\exp(-2B/T)$ lie on the unit circle, $|\tilde
z_j|=1$.\cite{LeeYang,YangLee} One of the corollaries of this
\emph{circle theorem} is that phase transitions at a fixed
temperature $T$ can only occur at zero magnetic field. A convenient
way to determine the positions of the Lee-Yang zeros in the complex
plane is to use a transfer matrix formulation. In this framework, the
occurrence of zeros in the partition function is equivalent to a
matching condition for the modulus of the two largest eigenvalues of
the transfer matrix, $|\Lambda_0(\tilde z)|=|\Lambda_1(\tilde
z)|$.\cite{KatsuraOhminami,YamadaNakano}

In 1965, Fisher suggested to study alternatively the zeros of the
canonical partition function in the complex inverse temperature
plane.\cite{FisherZeros} He showed that for the two-dimensional Ising
model without magnetic field, the Fisher zeros also lie on a unit
circle. Later, it was however realized that this is the exception
rather than the rule and that no statement equivalent to the Yang-Lee
circle theorem can be made.\cite{SaarloosKurtze} In general, the
Fisher zeros can, in addition to forming smooth curves, also densely
occupy entire regions of the complex plane. The fundamental difference
to the Lee-Yang study of zeros in the complex fugacity plane is that
the canonical partition function is not a simple polynomial in the
complex parameter $\beta$.

Many of these arguments can be transferred straightforwardly to a
study of the boundary partition function, Eq.~\eqref{Zz}. The Fisher
zeros $z_j$ will again determine the points where the potential
\begin{equation}
\label{fz}
f(z)=-\lim_{N\to\infty}\frac{1}{N}\ln Z(z)
\end{equation}
shows nonanalyticities. If such zeros close in on the imaginary axis,
$z_j\to\im t_j$, nonanalyticities will also occur in the real
time evolution of the quantum system and might be understood as
indicating a {\it dynamical phase transition}.\cite{HeylPolkovnikov}
However, it is important to note that $Z(\im t)$, contrary to
$Z_\textrm{th}$, is not a strictly positive function so that $Z(\im
t)=0$ is possible even for a finite system. An interpretation as a
dynamical phase transition has been supported by analytic results for
the one-dimensional transverse Ising model showing that the Fisher
zeros form lines in the complex plane which cross the imaginary axis
$z=\im t$ only if the quench leads across the equilibrium quantum
critical point.\cite{HeylPolkovnikov} Similarly, numerical
calculations for one-dimensional generalized Ising
models\cite{KarraschSchuricht} seem to suggest that the return rate
\begin{equation}
\label{lz}
l(t)=f(\im t)+f(-\im t)=-\lim_{N\to\infty}\frac{1}{N}\ln L(t)
\end{equation}
has nonanalyticities only for quenches across a quantum critical
point. As in the two-dimensional Ising case studied by Fisher one
might, however, wonder how general these results are given that the
boundary partition function $Z(z)$ is not a simple polynomial in $z$.

Even for integrable models that cannot be mapped onto free fermions
such as the transverse Ising model, it might be still possible to
investigate the Loschmidt echo analytically in certain cases by
considering the transfer matrix for the boundary partition function
$Z(z)$. Indeed, it has been recently shown that an analytic result for
the Loschmidt echo at imaginary times can be obtained for specific
quenches in the XXZ model by applying the Bethe ansatz to the
boundary transfer matrix.\cite{Pozsgay_LE} However, an analytic
continuation has so far only been possible for small times.
Furthermore, such an approach seems to be restricted to initial states
which are products of local two-site states.

Another recent approach is based on rewriting the Loschmidt echo as a
trace over a density matrix.\cite{Fagotti_LE} For a finite system, we
can write down a spectral representation of the Loschmidt amplitude
\begin{equation}
\label{Zz2}
Z(z)=\sum_n |\langle\Psi_0|n\rangle|^2 e^{-zE_n} = \Tr\{\rho_\textrm{diag} e^{-zH}\}\, ,
\end{equation}
where $\{|n\rangle\}$ are the eigenstates of $H$, $E_n$ the
corresponding eigenenergies, and the {\it diagonal ensemble} is
defined as
\begin{equation}
\label{diag_ens}
\rho_\textrm{diag}=\sum_n |\langle\Psi_0|n\rangle|^2 |n\rangle\langle n| \, .
\end{equation}
From the spectral representation \eqref{Zz2} one can easily obtain the
long-time average of the Loschmidt echo for finite systems
\begin{equation}
\label{Ltinf}
\overline L \equiv  \lim_{s\to\infty}\frac{1}{s}\int_0^s
L(t)\, dt\, =\, \sum_n |\langle\Psi_0|n\rangle|^4 \, . 
\end{equation}
Note that this is not simply the fidelity between the initial state
and the ground state of the final Hamiltonian, but rather involves a
sum over all eigenstates of the final Hamiltonian. In
Ref.~\onlinecite{Fagotti_LE}, the diagonal ensemble in \eqref{Zz2} was
replaced by the {\it generalized Gibbs ensemble} $\rho_\textrm{GGE}$
which is a function of all the local conservation
laws.\cite{RigolDunjkoPRL} For a nonintegrable system the generalized
Gibbs ensemble reduces to the canonical or grand canonical
ensemble.\cite{SirkerKonstantinidis} While it was shown in
Ref.~\onlinecite{Fagotti_LE} that the Loschmidt echo for the
transverse Ising model remains the same up to a factor $2$ if this
replacement is used, such a relation will not hold in general. The
'Loschmidt echo' with $\rho_\textrm{diag}\to \rho_\textrm{GGE}$ is a
different quantity which we will not consider here.

Our paper is organized as follows: In Sec.~\ref{TM} we describe the
similarities and differences between the thermal partition function
and the Loschmidt amplitude on the level of transfer matrices. Based
on this representation we formulate a numerical algorithm which allows
to calculate $Z(z)$ directly in the thermodynamic limit. Furthermore,
we have access to the spectrum of the transfer matrix and show that
nonanalyticities in $l(t)$ are a consequence of a crossing of the
leading with the next leading eigenvalue. In Sec.~\ref{XXZ} we give
the Hamiltonian of the one-dimensional XXZ model in a uniform and
staggered field which is the model we consider in the rest of the
paper. This model is integrable if the staggered field is zero and
nonintegrable otherwise. We recapitulate the phase diagram that
includes first and second orders, as well as
Berezinksy-Kosterlitz-Thouless transitions.  For specific quenches to
the free fermion point, the Fisher zeros and the Loschmidt echo can be
calculated analytically. We use the analytic solution in these cases
to discuss the structure of the Fisher zeros and its dependence on the
initial state. Furthermore, we use bosonization to obtain analytical
results for quenches within the Luttinger liquid phase and discuss the
applicability of this approximation by comparing with numerical data.
The results of both analytical approaches are presented in
Sec.~\ref{Analytics}. In all other cases we rely solely on numerical
calculations using the transfer matrix approach introduced in
Sec.~\ref{TM}. We present, in particular, a comparison between
quenches across different quantum critical lines in the XXZ model in
Sec.~\ref{Numerics} and give examples for nonanalytic behavior in
$l(t)$ for quenches which stay in the same phase. The general
conclusions which can be drawn from our analysis are discussed in the
final section.

\section{The transfer matrix approach}
\label{TM}
We will first briefly revisit the transfer matrix approach to
calculate the thermodynamics of one-dimensional quantum systems
discussing, in particular, the crossing of eigenvalues in the spectrum
of the transfer matrix. In a second step, we will extend this approach
to the boundary partition function $Z(z)$.

\subsection{The finite temperature formalism}
\begin{figure}
 \includegraphics*[width=1.0\columnwidth]{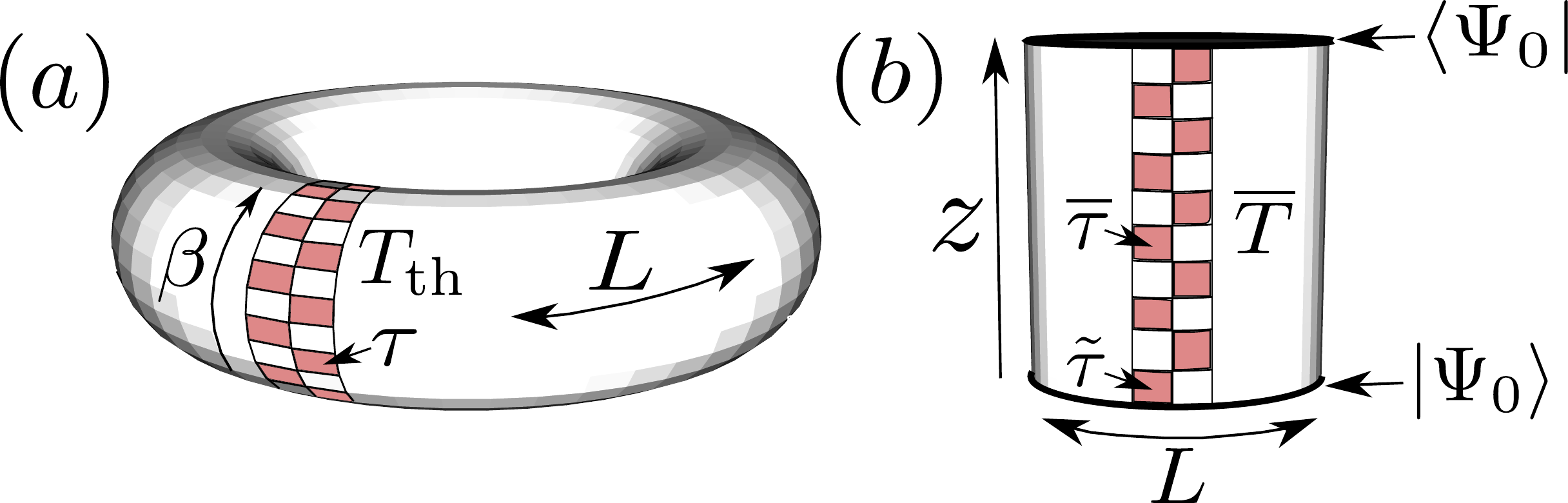}
 \caption{(Color online) (a) Torus geometry obtained after mapping a
   one-dimensional quantum system at finite temperatures with periodic
   boundary conditions to a two-dimensional classical model. (b)
   Cylinder geometry required to calculate the Loschmidt amplitude.
   The upper and lower ends of the cylinder are fixed by the initial
   state $|\Psi_0\rangle$.}
 \label{Fig1}
 \end{figure}

A one-dimensional quantum model at finite temperatures can be mapped
onto a two-dimensional classical model with the second dimension being
given by the inverse temperature $\beta$ (imaginary time). In
practice, such a mapping is achieved by a Trotter-Suzuki
decomposition of the partition function\cite{Trotter,Suzuki1,Suzuki1b}
\begin{equation}
\label{TS}
Z_\textrm{th}=\Tr\,\e^{-\beta H} =\lim_{M\to\infty} \Tr \left(\e^{-\delta_\beta H}\right)^M
\end{equation}
with $\beta=\delta_\beta\cdot M$ fixed. The infinitesimal evolution in
imaginary time can now be treated as a classical object and for a
Hamiltonian with nearest-neighbor interactions, $H=\sum_j h_{j,j+1}$,
the partition function becomes
\begin{equation}
\label{TS2}
Z_\textrm{th}=\lim_{M\to\infty} \Tr_L\left(\underbrace{\tau_{1,2}^{1,2}\tau_{2,3}^{2,3}\cdots\tau_{1,2}^{M-1,M}\tau_{2,3}^{M,1}}_{T_\textrm{th}}\right)^{L/2},
\end{equation}
where $L$ is the system size and we have assumed periodic boundary
conditions in the spatial direction (trace ``$\Tr_L$''). In the
imaginary time direction we have $M$ time slices and also periodic
boundary conditions. The geometry of the classical system is therefore
that of a torus as shown in Fig.~\ref{Fig1}(a). 

 The {\it quantum transfer matrix} $T_\textrm{th}$ is two columns wide
 and evolves along the spatial direction. It consists of Boltzmann
 weights
\begin{equation}
\label{tau}
\tau_{i,j}^{k,l}=\left[\exp(-\delta_\beta h_{i,j})\right]_k^l
\end{equation}
with $k,l\in\{1,\cdots,M\}$ indices labeling the sites of the
two-dimensional model in imaginary time direction, and
$i,j\in\{1,\cdots,L\}$ indices labeling the sites in the spatial
direction. We can now calculate the trace in the left and right
eigenbasis $\{\langle\phi^L_n|\}$ and $\{|\phi^R_n\rangle\}$ of the
$(M\times M)$ transfer matrix $T_\textrm{th}$ and find
\begin{equation}
\label{TS3}
Z_\textrm{th}= \lim_{M\to\infty}\left[ \Lambda_0^{L/2}\sum_{n=0}^{M-1} \left(\frac{\Lambda_n}{\Lambda_0}\right)^{L/2}\right] ,
\end{equation}
where $|\Lambda_0|\geq |\Lambda_1|\geq\cdots\geq|\Lambda_{M-1}|$ are
the eigenvalues of $T_\textrm{th}$ ordered in magnitude and we have
assumed that $L$ is even. For the following discussion, it is crucial
to note that the eigenvalues $\Lambda_i(\beta)$ will always be
continuous functions of inverse temperature $\beta$ if we follow their
evolution {\it without} reordering them according to their modulus
at every imaginary time step. For any finite system size $L$ the
partition function $Z_\textrm{th}$, as defined in \eqref{TS3}, will
therefore be an analytic function. However, in the thermodynamic
limit, $L\to\infty$, the partition function will be completely
determined by the largest eigenvalue $\Lambda_0$ giving rise to a free
energy
\begin{equation}
\label{freeE}
f_\textrm{th}=-\lim_{L\to\infty}\frac{T}{L}\ln Z_\textrm{th}=-\frac{T}{2}\ln\Lambda_0 \, .
\end{equation}
Since eigenvalues can cross, this opens up the possibility for
nonanalyticities. To understand the physical meaning of such
crossings in the spectrum of the transfer matrix, one has to consider
two-point correlation functions between two operators $A$ and $B$
given at distances $r\geq 2$ by
\begin{eqnarray}
\label{CF}
&& \langle A_1 B_{r+1}\rangle = \Tr\{A_1B_{r+1}\e^{-\beta H}\}/Z_\textrm{th} \\
&=& \Tr\{ T_\textrm{th}(A_1)T_\textrm{th}^{[r/2]-1}T_\textrm{th}(B_{r+1})T_\textrm{th}^{L/2-[r/2]-1}\}/\Tr T_\textrm{th}^{L/2} \nonumber \\
&\stackrel{L\to\infty}{\to}& \langle\phi_0^L|T_\textrm{th}(A_1)T_\textrm{th}^{[r/2]-1}T_\textrm{th}(B_{r+1})|\phi_0^R\rangle/\Lambda_0^{[r/2]+1} \nonumber \\
&=& \sum_n \frac{\langle\phi_0^L|T_\textrm{th}(A_1)|\phi_n^R\rangle\langle\phi_n^L|T_\textrm{th}(B_{r+1})|\phi_0^R\rangle}{\Lambda_0\Lambda_n}\left(\frac{\Lambda_n}{\Lambda_0}\right)^{[r/2]} \, ,\nonumber
\end{eqnarray} 
where $[r/2]$ denotes the largest integer smaller than $r/2$.
$T_\textrm{th}(A)$ and $T_\textrm{th}(B)$ are transfer matrices as
defined in \eqref{TS2} with the operator $A$ or $B$ inserted at the
appropriate position.  The correlation function therefore has the
asymptotic expansion
\begin{eqnarray}
\label{CF2}
\langle A_1 B_{r+1}\rangle &=& \frac{\langle\phi_0^L|T_\textrm{th}(A_1)|\phi_0^R\rangle\langle\phi_0^L|T_\textrm{th}(B_{r+1})|\phi_0^R\rangle}{\Lambda_0^2}\nonumber \\
&+& \sum_{n\geq 1} M_n \e^{ik_nr}\e^{-r/\xi_n},
\end{eqnarray} 
with matrix elements
\begin{equation}
\label{Mn}
M_n= \frac{\langle\phi_0^L|T_\textrm{th}(A_1)|\phi_n^R\rangle\langle\phi_n^L|T_\textrm{th}(B_{r+1})|\phi_0^R\rangle}{\Lambda_0\Lambda_n} \, ,
\end{equation}
wave vectors
\begin{equation}
\label{kn}
k_n=\frac{1}{2}\arg\left(\frac{\Lambda_n}{\Lambda_0}\right)+l\pi \quad (l=0\; \mbox{or}\; l=1) \, ,
\end{equation}
and correlation lengths
\begin{equation}
\label{xin}
\xi_n^{-1}=\frac{1}{2}\ln\left|\frac{\Lambda_0}{\Lambda_n}\right| \, .
\end{equation}
A crossing of a next-leading eigenvalue $\Lambda_n$ with the leading
eigenvalue $\Lambda_0$ thus implies a diverging correlation length
$\xi_n$. For one-dimensional quantum systems there are no finite
temperature phase transitions so that this will never happen and a gap
between the leading and next-leading eigenvalue will persist for any
finite temperature. A divergence of correlation lengths will only
occur in the limit $T\to 0$ when approaching a quantum critical point
or line.  However, crossovers in the spectrum between eigenvalues
$\Lambda_i$ and $\Lambda_j$ with $i,j>0$ are possible and lead to 
nonanalyticities in the correlation lengths. Such
nonanalyticities have been analyzed in detail using the quantum
matrix approach and Bethe ansatz for the integrable XXZ
model\cite{KluemperScheeren} and by numerical density-matrix
renormalization group methods for transfer matrices
(TMRG)\cite{BursillXiang,WangXiang,Shibata} for the nonintegrable
$t-J$ chain\cite{SirkerKluemperEPL,SirkerKluemperPRB}, spin-orbital
models\cite{SirkerKhaliullin,SirkerSu4,SirkerDamerau}, and an extended
Hubbard model.\cite{GlockeSirker}

\subsection{The Loschmidt echo}
The finite temperature formalism described above can be
straightforwardly generalized to a calculation of the Loschmidt
amplitude $Z(z)$ defined in Eq.~\eqref{Zz}.  There are only two
modifications: (1) Instead of performing a Trotter-Suzuki
decomposition for the imaginary time evolution operator $\e^{-\beta
  H}$ we now have to do such a decomposition for the evolution
operator $\e^{-zH}$ with an arbitrary complex number $z$. (2) The
geometry is different. Instead of a torus we now have to consider a
cylinder with the boundaries in the $z$-direction fixed by the initial
state $|\Psi_0\rangle$ as shown in Fig.~\ref{Fig1}(b).

In complete analogy to Eq.~\eqref{TS2} we obtain
\begin{equation}
\label{LE1}
Z(z)=\lim_{M\to\infty}\Tr_L \overline T^{L/2} \, ,
\end{equation}
where the trace is taken along the spatial direction due to the
assumed periodic boundary conditions. The quantum transfer matrix
$\overline T$ now has boundary degrees of freedom fixed by the initial
configuration $|\Psi_0\rangle$,
\begin{equation}
\label{LE2}
\overline T = \widetilde\tau_{1,2}^{1,2}\overline\tau_{2,3}^{2,3}\cdots\overline\tau_{1,2}^{M-1,M}\widetilde\tau_{2,3}^{M,M+1} \, ,
\end{equation}
with weights $\overline\tau_{i,j}^{k,l}$ as given in Eq.~\eqref{tau}
with $\delta_\beta\to \delta_z$. The first and last weight $\widetilde\tau$ are, however, modified with matrixelements
\begin{equation}
\label{tau2}
\widetilde\tau_{1,2}^{1,2}=\langle s^1_1 s^1_2| \exp(-\delta_z h_{1,2}) | s^2_1 s^2_2 \rangle \, ,
\end{equation}
where the degrees of freedom $s^1_1$ and $s^1_2$ are fixed by the
initial state $|\Psi_0\rangle$. $\widetilde\tau_{2,3}^{M,M+1}$ is
defined analogously.

For the boundary partition function $Z(z)$ the formula \eqref{TS3}
thus applies as well with the eigenvalues $\Lambda_i$ of the thermal
transfer matrix $T_\textrm{th}$ replaced by the eigenvalues
$\overline\Lambda_i$ of the boundary transfer matrix $\overline T$. If
we order these eigenvalues again by magnitude,
$|\overline\Lambda_0|\geq\overline|\Lambda_1|\geq\cdots\geq|\overline\Lambda_{M-1}|$,
then the return rate $f(z)$ defined in Eq.~\eqref{fz} is given in the
thermodynamic limit by
\begin{equation}
\label{fz2}
f(z)=-\frac{1}{2}\lim_{M\to\infty}\ln\overline\Lambda_0 \, .
\end{equation}
However, there is now no longer necessarily a gap between the largest
and next-leading eigenvalues. Crossings are possible which can make
$\overline\Lambda_0(z)$ and thus $f(z)$ a nonanalytic function. In
order to find the lines where the Fisher zeros accumulate in the
thermodynamic limit we therefore have to find the $z_0$ values for
which
\begin{equation}
\label{Fisher_cond}
|\overline\Lambda_0(z_0)| = |\overline\Lambda_k(z_0)|\;\,\mbox{and}\;\, |\overline\Lambda_0(z_0\pm\delta)| \neq |\overline\Lambda_k(z_0\pm \delta)|
\end{equation}
with $|\delta|\ll 1$. In this case, we have two regions where either
$\overline\Lambda_0$ or $\overline\Lambda_k$ dominate with the
matching condition \eqref{Fisher_cond} leading to a nonanalyticity in
$f(z)$. Note that in both regions the largest eigenvalue can be
degenerate. Only a true crossing---not a degeneracy---leads to
nonanalytic behavior.  Alternatively, we can understand the crossing
of two eigenvalues as a divergence of a correlation length
\begin{equation}
\label{corr_LE}
\overline\xi_k^{-1}=\frac{1}{2}\ln\left|\frac{\overline\Lambda_0}{\overline\Lambda_k}\right| \, .
\end{equation}
In this sense, the crossing is indeed indicating a dynamical phase
transition.
Finally, we want to explicitly give the formula for the return
amplitude in the thermodynamic limit,
\begin{equation}
\label{lt}
l(t)=f(\im t)+f(-\im t)=-\frac{1}{2}\ln |\overline\Lambda_0|^2 .
\end{equation}

In the next section we describe a numerical algorithm to calculate
the boundary transfer matrix $\overline T$. Using
Eqs.~\eqref{Fisher_cond} and \eqref{lt} we can then determine
numerically the Fisher zeros and the return amplitude for any
translationally invariant one-dimensional model with short-range
interactions directly in the thermodynamic limit.

\subsection{The numerical implementation}
\label{implementation}
For the numerical simulations, we use the light cone renormalization
group (LCRG) algorithm introduced in Ref. \onlinecite{EnssSirker}.
Working in the thermodynamic limit, we have direct access to the
transfer matrix $\overline T$, from which we can calculate the largest
eigenvalues $\overline\Lambda_n$ using an iterative Lanczos
eigensolver. The LCRG algorithm was originally introduced to calculate
expectation values of observables $O$ following a quantum quench,
$\langle O(t)\rangle=\langle\Psi_0|\e^{iHt}O\e^{-iHt}|\Psi_0\rangle$.
In this case the corresponding two-dimensional classical model has the
form of a light cone with Boltzmann weights of the forward and
backward time evolution canceling outside of the light cone which has
the operator $O$ at its center. No such structure exists for the
Loschmidt amplitude where only the evolution operator forward in time
is present. The transfer matrix in the LCRG algorithm, which evolves
along the horizon of the light cone, can nevertheless still be used to
construct a two-dimensional lattice which is infinite in the spatial
direction and the relation \eqref{fz2} remains valid. Alternatively,
one can use an infinite time evolving block decimation (iTEBD)
algorithm\cite{VidaliTEBD} to construct the transfer matrix $\overline
T$. This method has been used to calculate $l(t)$ in the thermodynamic
limit for generalized Ising models in Ref.
\onlinecite{KarraschSchuricht}.  However, in this work only the
real-time evolution has been considered while we will use a general
complex parameter $z$ in the evolution operator which gives us, in
addition, also access to the Fisher zeros in the complex plane.

The ground state of the initial Hamiltonian is obtained in our
algorithm by applying an imaginary time evolution $e^{-\beta H}$ to an
arbitrary state. For gapped initial Hamiltonians, we have $e^{-\beta
  H}=|\Psi_0\rangle\langle\Psi_0|+\mathcal{O}(e^{-\beta\Delta})$,
with $\Delta=E_1-E_0$ the energy gap. In this case the convergence is
fast once imaginary times $\beta\gg 1/\Delta$ are reached. For an initial
state in a gapless phase, this projection method is much more
problematic and can be the source of significant numerical
errors.\cite{Schollwock_review} In the integrable case, we will check
the convergence by comparing with the exact ground-state
energy.\cite{tak99} The projection onto the ground state is followed
by the standard time evolution with $U(z)=e^{-zH}$, where $z=R+\im t$
is a complex parameter.  All simulations are carried out using an
adaptive block dimension of up to $\chi=768$ states such that the
discarded weight at each step remains smaller than $10^{-15}$. We can,
in addition, get an upper bound on the possible simulation time
$t_\textrm{max}$ by calculating the entanglement entropy
$S_\textrm{ent}$. For a block dimension $\chi$, the maximal
entanglement entropy which can be represented is given by
$S^\textrm{max}_\textrm{ent}=\ln\chi$. Since $S_\textrm{ent}(t)=c t$
with a constant $c$, we can easily determine $t_\textrm{max}$ by the
condition $ct_\textrm{max}=S^\textrm{max}_\textrm{ent}$ and we make
sure that in our simulations $t$ is always much smaller than
$t_\textrm{max}$.

\begin{figure}
 \includegraphics*[width=1.0\columnwidth]{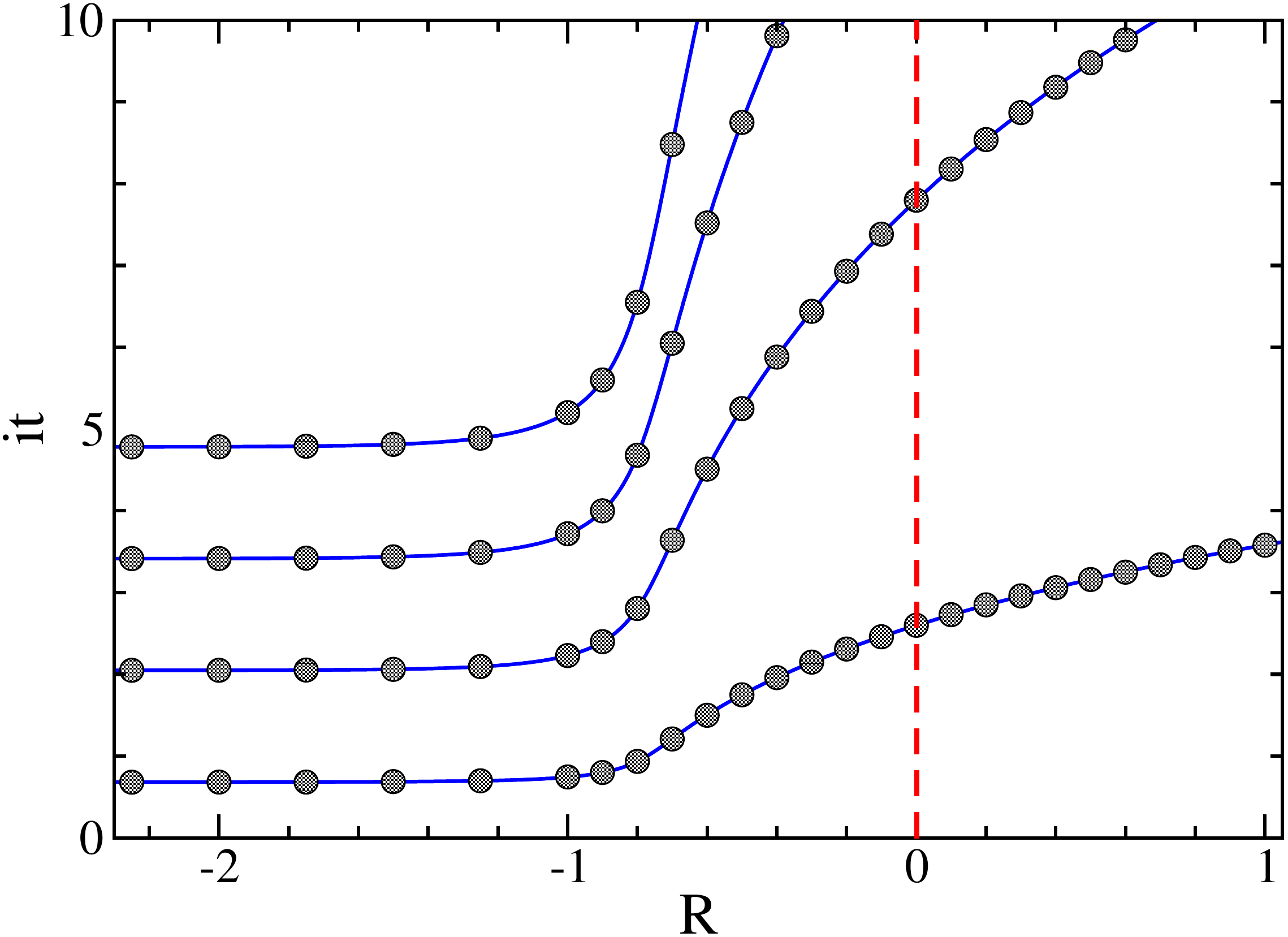}
 \caption{Lines of Fisher zeros in the complex plane, $z=R+it$, for a
   quench $g=0.4\to 1.3$ in the transverse Ising model
   \eqref{trans_Ising}. The exact results\cite{HeylPolkovnikov}
   (lines) are compared with numerical scans through the complex plane
   using the LCRG algorithm (circles).}
 \label{Fig_Ising}
 \end{figure}

To show that it is indeed possible to determine the lines of Fisher
zeros using the condition \eqref{Fisher_cond} and the LCRG algorithm,
we consider the one-dimensional transverse Ising model
\begin{equation}
\label{trans_Ising}
H=-\frac{1}{2}\sum_i \sigma^z_i\sigma^z_{i+1} +\frac{g}{2}\sum_i \sigma^x_i
\end{equation}
where an analytical solution can be
obtained.\cite{QuanSong,HeylPolkovnikov} In Fig.~\ref{Fig_Ising}, the
analytical result taken from Ref.~\onlinecite{HeylPolkovnikov} is
compared with numerical data obtained by the LCRG algorithm for a
quench across the quantum phase transition at $g=1$, showing excellent
agreement.

\section{The model}
\label{XXZ}
\begin{figure}
 \includegraphics*[width=1.0\columnwidth]{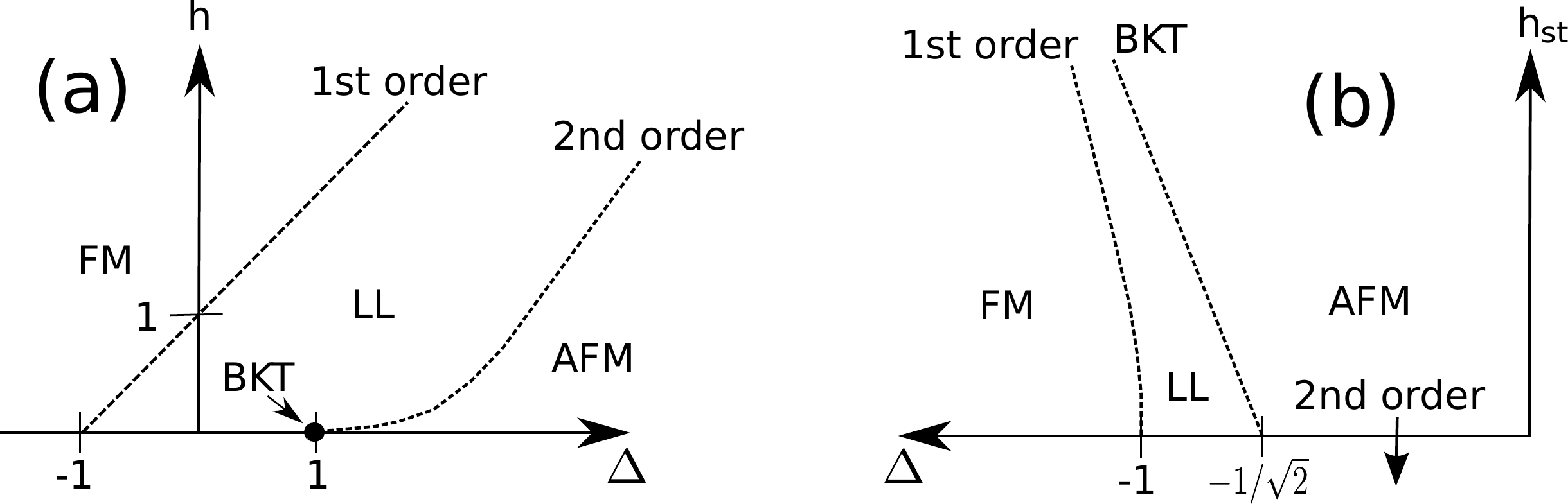}
 \caption{Schematic phase diagrams for the Hamiltonian \eqref{model}
   for (a) the integrable case $h_\textrm{st}=0$, and (b) the
   nonintegrable case with staggered field but $h=0$. The phase
   diagrams consist of gapped ferromagnetic (FM) and antiferromagnetic
   (AFM) phases as well as a gapless Luttinger liquid (LL) phase. In
   (a), the transition at $\Delta=1$ and $h=0$ is of
   Berezinsky-Kosterlitz-Thouless (BKT) type, while in (b), the whole
   line between AFM and LL phase is a BKT transition.}
 \label{Fig_phase_diag}
 \end{figure}

In order to exemplarily investigate the Fisher zeros, the
nonanalyticities in the Loschmidt echo, and their relation to
equilibrium quantum phase transitions, we consider quenches in a
spin-$1/2$ XXZ model given by
\begin{eqnarray}
\label{model}
H&=&J\sum_j\left\{ \frac{1}{2}[S^+_jS^-_{j+1}+S^+_{j+1}S^-_j] +\Delta S^z_jS^z_{j+1} \right\} \nonumber \\
&+& \sum_j \left( h +(-1)^j h_\textrm{st}\right)S^z_j \, .
\end{eqnarray}
Here $J$ is the exchange coupling constant, $\Delta$ parametrizes the
exchange anisotropy, $h$ is a uniform, and $h_\textrm{st}$ is a staggered
magnetic field. We will concentrate on investigating the model for the
following two cases: (I) $h_\textrm{st}=0$. In this case, the model is
integrable by Bethe ansatz. The phase diagram, shown schematically in
Fig.~\ref{Fig_phase_diag}(a), consists of a ferromagnetic (FM), a
Luttinger liquid (LL), and an antiferromagnetic (AFM)
phase.\cite{tak99} The phase transitions are of first, second, and
Berezinsky-Kosterlitz-Thouless (BKT) type as indicated in
Fig.~\ref{Fig_phase_diag}(a). (II) $h=0$ and $h_\textrm{st}\neq 0$. In
this case, the model is not integrable. Again, we have FM, LL, and AFM
phases, separated by first-order, second-order, and BKT transitions as
indicated in the schematic phase diagram in
Fig.~\ref{Fig_phase_diag}(b).\cite{AlcarazMalvezzi,OkamotoNomura_st_field}

 It is clear right from the start that the Loschmidt echo is trivial
 for certain quenches across those phase transitions. First, this is
 always the case for quenches starting in the FM phase because the
 ferromagnetic state is always an eigenstate of the Hamiltonian
 \eqref{model}. No matter what parameters we choose for the final
 Hamiltonian, we will always obtain $|Z(z)|\equiv 1$. Furthermore,
 $|Z(z)|$ is also always equal to one if we only quench the uniform
 magnetic field, $h_\textrm{ini}\to h_\textrm{fin}$, because
 $[H,\sum_j S^z_j]=0$. There is thus no direct correspondence between
 the zero temperature phase transitions and nonanalyticities in the
 Loschmidt echo in general.  In Sec.~\ref{Numerics} we will even show
 that nonanalyticities in the Loschmidt echo can occur for this model
 for quenches which do not cross any of the transition lines. In the
 following, we will denote the quench from a given set of initial to
 some final microscopic parameters by
 $(\Delta_\textrm{ini},h_\textrm{ini},h_\textrm{st,ini})\to
 (\Delta_\textrm{fin},h_\textrm{fin},h_\textrm{st,fin})$ and set
 $J=1$.

\section{Analytical results for bilinear Hamiltonians}
\label{Analytics}
Before numerically investigating the model \eqref{model}, first we want to
discuss cases where we can exactly calculate $Z(z)$ or where we
might be able to approximate the exact time evolution by that of a
bilinear Hamiltonian obtained by bosonization.

\subsection{Free fermion case}
\label{FreeFermions}
For $\Delta=0$, we can map the Hamiltonian \eqref{model} to a free
fermion Hamiltonian using the Jordan-Wigner transformation
\begin{equation}
\label{JW}
S^+_j=(-1)^jc_j^\dagger \e^{i\pi\phi_j},\, S^-=(-1)^j c_j \e^{-i\pi\phi_j},\, S^z_j=n_j-1/2 
\end{equation}
with $n_j=c^\dagger_j c_j= S^+_jS^-_j$, $\phi_j=\sum_{l=1}^{j-1} n_l$,
where $c_j$ ($c_j^\dagger$) are annihilation (creation) operators of
spinless fermions at site $j$. In the following we assume that the
spin chain has an even number of lattice sites $N$ and periodic
boundary conditions. Equation \eqref{JW} then leads to the following
fermionic Hamiltonian
\begin{eqnarray}
\label{fF1}
H_0&=&-\frac{J}{2}\left[\sum_{j=1}^{N-1} (c^\dagger_jc_{j+1}+h.c.) -\e^{i\pi M} (c_N^\dagger c_1+h.c.)\right] \nonumber \\
&+&\sum_{j=1}^N [h+(-1)^jh_\textrm{st}](n_j-1/2)
\end{eqnarray}
where $M=\sum_{j=1}^N n_j$ is the total particle number. Note that the
boundary conditions for the fermions are only periodic if $M$ is odd.
Otherwise there is a phase shift.\cite{LiebSchultzMattis} In order to
diagonalize $H_0$ we can define a Fourier transform as
\begin{equation}
\label{FT}
c_j=\frac{1}{\sqrt{N}}\sum_{n=1}^N \e^{-ij(k_n+a)}
\end{equation}
where $k_n=2\pi n/N$, $a = \pi/N$ for $M$ even, and $a=0$ for $M$ odd,
leading to
\begin{eqnarray}
\label{fF2}
H_0&=&\sum_{n=1}^N (\varepsilon_{k_n}-h) c_{k_n}^\dagger c_{k_n} - h_\textrm{st}\sum_{n=1}^N c_{k_n}^\dagger c_{k_{n+N/2}} \nonumber \\
&=&\sum_{n=1}^{N/2} [(\varepsilon_{k_n}-h) c_{k_n}^\dagger c_{k_n}-(\varepsilon_{k_n}+h) d_{k_n}^\dagger d_{k_n}]\nonumber \\
&-& h_\textrm{st}\sum_{n=1}^{N/2} [c_{k_n}^\dagger d_{k_n}+d_{k_n}^\dagger c_{k_n}] 
\end{eqnarray}
with $\varepsilon_{k_n}=-J\cos (k_n+a)$ . In the second and third lines,
we have introduced new operators $d_{k_n}$ by folding the band back
into a reduced Brillouin zone. This bilinear Hamiltonian can now
easily be diagonalized by a rotation
\begin{eqnarray}
\label{fF3}
H_0 &=&\sum_{n=1}^{N/2} \left[\left(\sqrt{\varepsilon_{k_n}^2+h_\textrm{st}^2}-h\right)\alpha_{k_n}^\dagger \alpha_{k_n}\right. \\
 &-& \left.\left(\sqrt{\varepsilon_{k_n}^2+h_\textrm{st}^2}+h\right)\beta_{k_n}^\dagger \beta_{k_n} \right]. \nonumber
\end{eqnarray}
Note that the rotation needed to diagonalize the Hamiltonian is a
function of the microscopic parameters $J,\, h,\, h_\textrm{st}$.

\begin{figure}
 \includegraphics*[width=1.0\columnwidth]{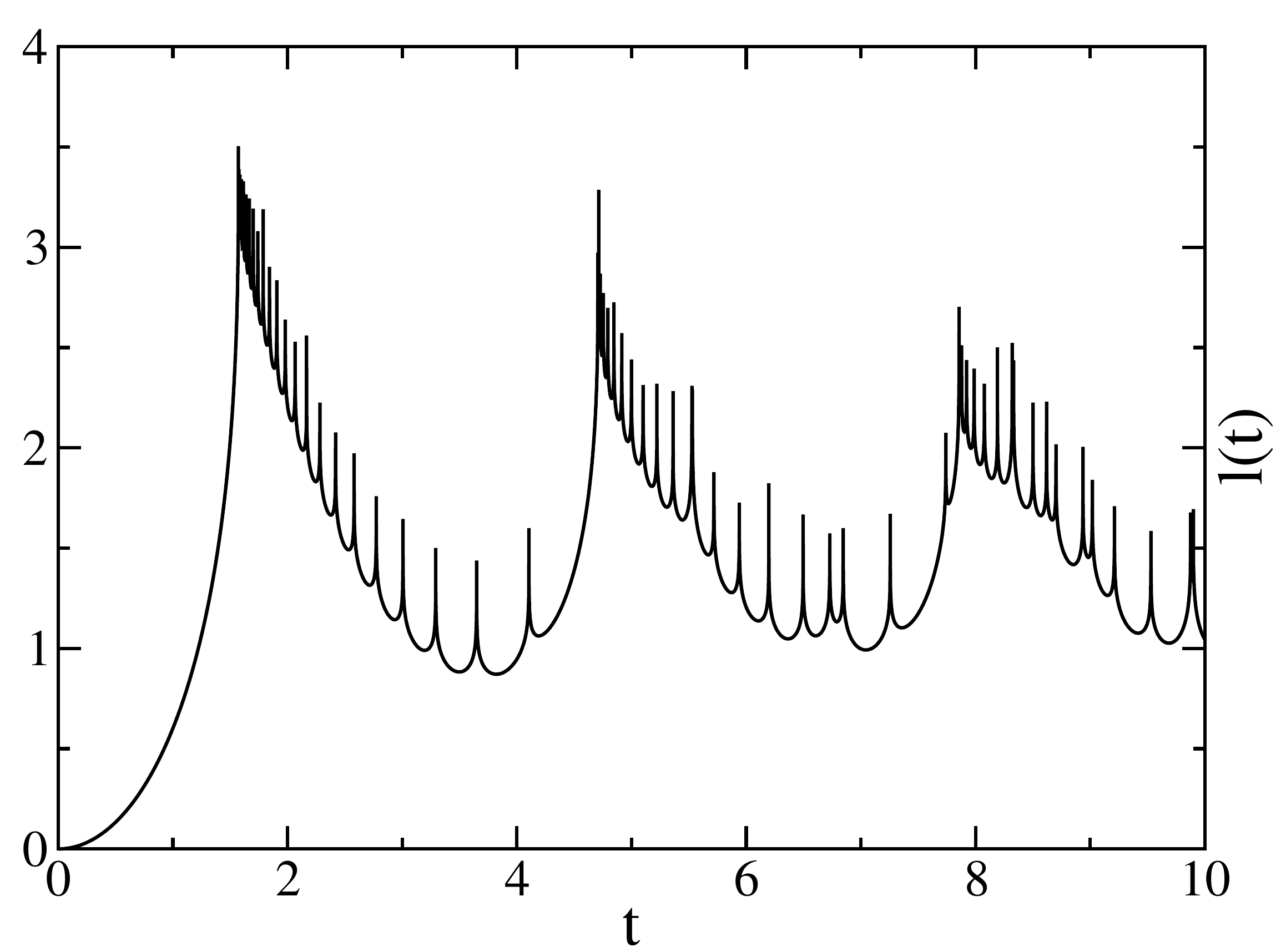}
 \caption{(Color online) The return rate $l(t)$, Eq.~\eqref{fF4}, for
   a system with $N=120$ sites. $l(t)$ diverges at times $t_c$ given
   by Eq.~\eqref{fF_tc_finite}.}
 \label{Fig_fF_N120}
 \end{figure}

\begin{figure}
 \includegraphics*[width=1.0\columnwidth]{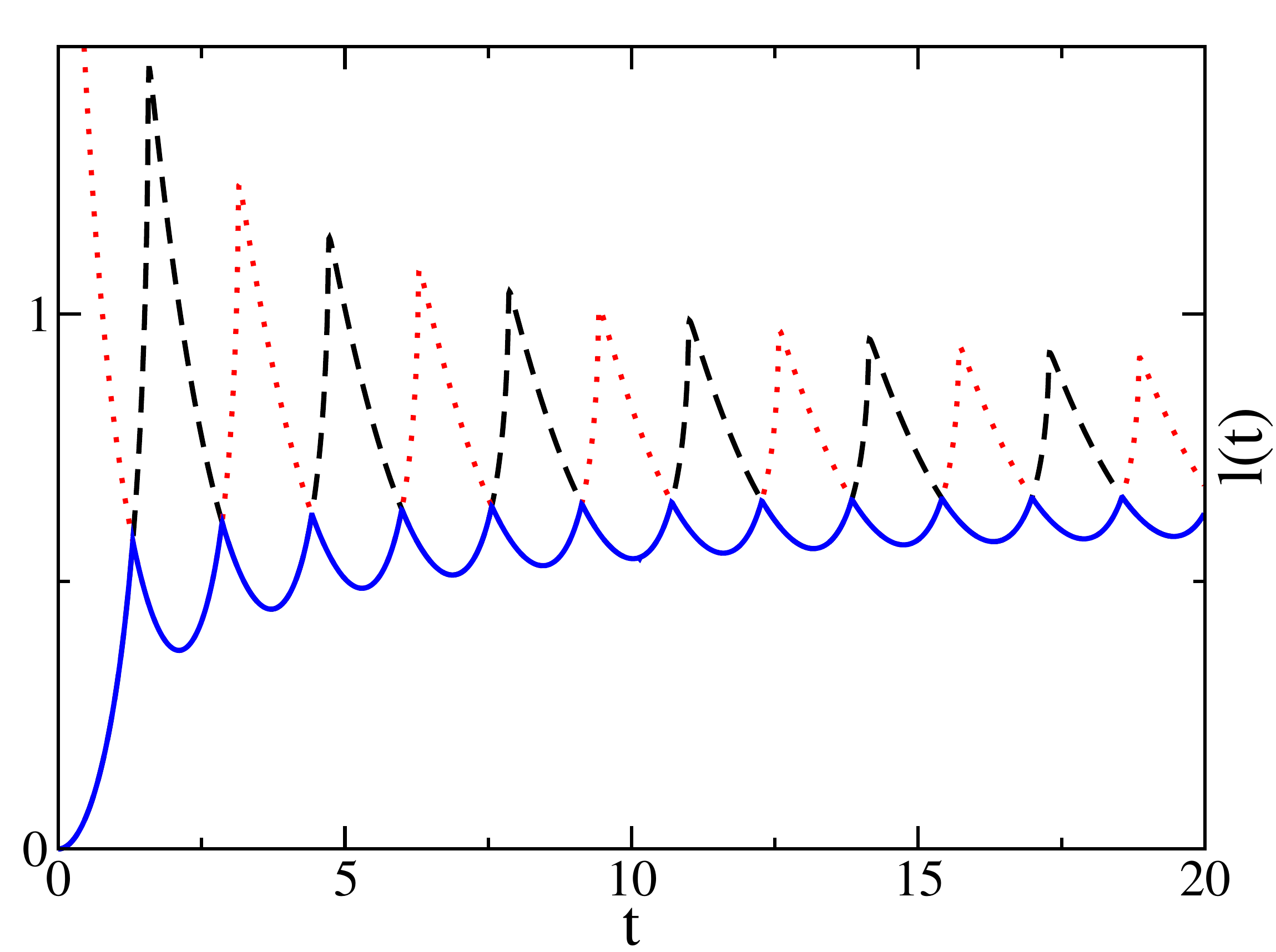}
 \caption{(Color online) The return rate $l(t)$ in the thermodynamic
   limit, Eq.~\eqref{fF4} (dashed line), the two contributions
   contributions in Eq.~\eqref{Neel_symm3} shown separately (dashed
   and dotted lines), and the minimum of these two contributions
   (solid line).}
 \label{Fig_fF}
 \end{figure}

Now we can consider various quenches in the free fermion model
\eqref{fF3} by changing these parameters. We start with a particularly
simple case where the initial state $|\Psi_0\rangle$ is the ground
state of $H_0$ with $J=h=0$ and $h_\textrm{st}\neq 0$. This is nothing
but a N\'eel state that we can express as $|\Psi_0\rangle = \prod_n
\beta^\dagger_{k_n} |0\rangle $ using the operators defined in
\eqref{fF3}. As the final Hamiltonian we take the Hamiltonian in
\eqref{fF2} for $h=h_\textrm{st}=0$, which is already diagonal. In this
case the relation between the old and new operators is particularly
simple:
\begin{equation}
c_{k_n} = \frac{1}{\sqrt{2}}(\alpha_{k_n}+\beta_{k_n}),\quad d_{k_n} = \frac{1}{\sqrt{2}}(\alpha_{k_n}-\beta_{k_n}).
\end{equation}
The time evolution of the initial state is now trivially given by
\begin{equation}
  \e^{-zH}|\Psi_0\rangle =\frac{1}{\sqrt{2}}\prod_{n=1}^{N/2}\left[\left(\e^{-z\varepsilon_{k_n}} c_{k_n}^\dagger - \e^{z\varepsilon_{k_n}}d_{k_n}^\dagger\right)\right]|0\rangle
\end{equation}
and the Loschmidt amplitude becomes $Z(z)=\prod_n \cosh
(z\varepsilon_{k_n})$. For the return rate this yields
\begin{eqnarray}
\label{fF4}
l(t)&=&-\frac{1}{N}\sum_{n=1}^{N/2}\ln \cos^2 (t\varepsilon_{k_n}) \\
&\stackrel{N\to\infty}{\to}& -\frac{1}{2\pi}\int_{-\pi/2}^{\pi/2} dk\,\ln \cos^2 (t\varepsilon_k). \nonumber
\end{eqnarray}
Here we want to emphasize again an important difference to the thermal
partition function. While $Z_\textrm{th}(z)$ can never have zeros on
the real axis, the Loschmidt amplitude $Z(z)$ can become zero for
$z=\im t$. This simply means that the time evolved state has become
orthogonal to the initial state. From \eqref{fF4}, we see that this
indeed happens for the considered quench and finite system size $N$ at
times
\begin{equation}
\label{fF_tc_finite}
t_c =\frac{\pi}{2\varepsilon_{k_n}}(2m+1),\quad m\in\mathbb{Z},\; n=0,\cdots, N-1.
\end{equation}
At these times, the return rate $l(t)$ diverges, see
Fig.~\ref{Fig_fF_N120}. 
 In the thermodynamic limit, however, only cusps in the return rate
 $l(t)$ remain. The integral in \eqref{fF4} is dominated by
 contributions from the band edge, $k\approx 0$, and cusps occur if
 $\cos(t\cos(k=0))=0$, i.e., at times
\begin{equation}
\label{fF_tc}
t_c =\frac{\pi}{2}(2m+1)
\end{equation}
as shown in Fig.~\ref{Fig_fF} (dashed line).

\begin{figure}
 \includegraphics*[width=1.0\columnwidth]{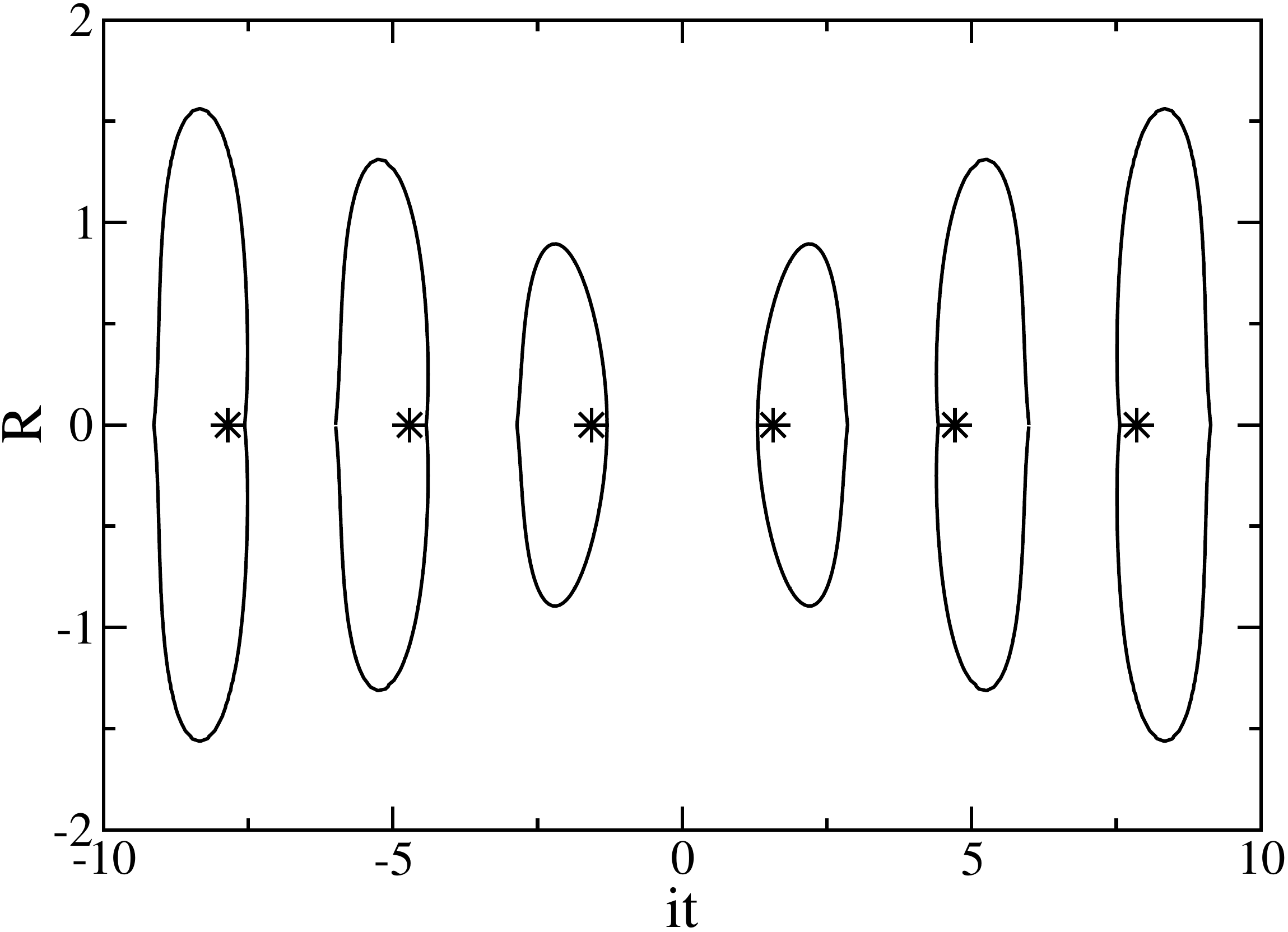}
 \caption{Lines of Fisher zeros obtained by Eq.~\eqref{Neel_symm4} for
   the quench from the symmetric N\'eel state \eqref{Neel_symm} to
   the free fermion point (solid lines), and Fisher zeros at times
   $t_c$, Eq.~\eqref{fF_tc}, for the quench from the polarized state
   (symbols).}
 \label{Fig_fF2}
 \end{figure}

 The location of the Fisher zeros depends, however, very sensitively
 on the initial state. For the N\'eel state chosen above the zeros all
 lie on the real time axis. This changes completely if we consider the
 symmetric combination of the two possible N\'eel states, i.e.,
\begin{equation}
\label{Neel_symm}
|\Psi_0\rangle =\frac{1}{\sqrt{2}}\left(\prod_n \beta_{k_n}^\dagger +\prod_n\alpha_{k_n}^\dagger\right) |0\rangle \, .
\end{equation}
In this case the Loschmidt amplitude becomes
\begin{equation}
\label{Neel_symm2}
Z(z)=\prod_n \cosh(z\varepsilon_{k_n}) + \prod_n\sinh(z\varepsilon_{k_n}) \, .
\end{equation}
In the thermodynamic limit, always one of the two contributions
dominate depending on the parameter $z$. The return rate is, in
particular, given by
\begin{eqnarray}
\label{Neel_symm3}
l(t)=-\frac{1}{2\pi}\min &&\!\!\!\!\!\left\{ \int_{-\pi/2}^{\pi/2} dk\, \ln\cos^2(t\varepsilon_k) \right. ,\\
 &&\left.\int_{-\pi/2}^{\pi/2} dk\, \ln\sin^2(t\varepsilon_k)\right\} \nonumber
\end{eqnarray}
and is shown in Fig.~\ref{Fig_fF} (solid line). Fisher zeros in the
complex plane occur if the two contributions in \eqref{Neel_symm3},
taken as function of the complex parameter $z=R+\im t$, are of equal
magnitude and are therefore determined by
\begin{equation}
\label{Neel_symm4}
0=\int_0^{\pi/2} dk\, \ln |\tanh(z\varepsilon_k)|^2 \, .
\end{equation}
A plot of the Fisher zeros for this quench, obtained by numerically
solving \eqref{Neel_symm4}, is shown in Fig.~\ref{Fig_fF2}.
 The separate Fisher zeros at real times $t_c$ given by
 Eq.~\eqref{fF_tc} for the quench from the N\'eel state thus turn into
 closed lines in the complex plane if we instead quench from the
 superposition of N\'eel states, Eq.~\eqref{Neel_symm}.

\subsection{Bosonization in the Luttinger liquid phase}
\label{Bosonization}
A one-dimensional quantum system with gapless excitations can be
described at low energies as a Luttinger liquid\cite{GiamarchiBook}
\begin{eqnarray}
\label{Bos1}
H_b&=&v_F\sum_{q>0} q \left\{\left(1+\frac{g_4}{2\pi v_F}\right)\left[ b_{Rq}^\dagger b_{Rq} + b_{Lq}^\dagger b_{Lq}\right]\right. \nonumber \\
&+& \left. \frac{g_2}{2\pi v_F} \left[ b_{Rq}^\dagger b^\dagger_{Lq} + b_{Lq}b_{Rq}\right]\right\} \, . 
\end{eqnarray}
Here $v_F$ is the Fermi velocity, and the bosonic operators
$b_{R/L,q}$ describe the right/left moving collective excitations with
momentum $q$. $g_{2,4}$ are forward scattering amplitudes. 

We set $\bar{g}_i=g_i/(2\pi v_F)$ and consider a quench where the
scattering amplitudes are changed from values $\bar{g}_{i,1}$ to $\bar
g_{i,2}$ in a quench. For both sets of parameters the Hamiltonian
\eqref{Bos1} is bilinear and can be diagonalized by a Bogoliubov
transform:
\begin{equation}
\label{Bos2}
\alpha_{iq}^{R/L}= u_i b_{R/L,q} + v_i b_{L/R,q}^\dagger 
\end{equation}
with parameters
\begin{equation}
\label{Bos3}
v_i^2=\frac{1}{2}\left[\frac{1+\bar g_{4,i}}{\sqrt{(1+\bar g_{4,i})^2-\bar g_{2,i}^2}}+1\right] =\frac{(1+K_i)^2}{4K_i},
\end{equation}
$u_i^2=v_i^2-1$, and $i=1,2$. In Eq.~\eqref{Bos3} we have also
introduced the Luttinger parameters $K_i$.

The two new operator sets $\alpha_{1q}^{R/L}$ and $\alpha_{2q}^{R/L}$
are then again related by a Bogoliubov transform,
\begin{eqnarray}
\label{Bos4}
\alpha_{1q}^{L\dagger} &=& (v_1u_2-u_1v_2)\alpha_{2q}^R + (u_1u_2-v_1v_2)\alpha_{2q}^{L\dagger} \nonumber \\
\alpha_{1q}^{R} &=& (\underbrace{u_1v_2-v_1u_2}_{v'})\alpha_{2q}^{L\dagger} + (\underbrace{u_1u_2-v_1v_2}_{u'})\alpha_{2q}^{R}\, . \nonumber \\
\end{eqnarray}
Now it is straightforward to calculate the Loschmidt amplitude 
\begin{eqnarray}
\label{Bos5}
Z(z)&=& _{\alpha_1}\langle 0| e^{-zH_2}|0\rangle_{\alpha_1}/ _{\alpha_1}\langle 0|0\rangle_{\alpha_1} \\
&=& _{\alpha_2}\langle 0|T^\dagger e^{-zH_2}T|0\rangle_{\alpha_2}/_{\alpha_2}\langle 0|T^\dagger T|0\rangle_{\alpha_2}  \nonumber
\end{eqnarray}
with
\begin{equation}
\label{Bos52}
T=\exp\left\{\frac{v'}{c'}\sum_q \alpha_{2q}^{R\dagger} \alpha_{2q}^{L\dagger }\right\}
\end{equation}
where $|0\rangle_{\alpha_1}$ is the ground state before the quench and $H_2$ the Hamiltonian after the quench given by
\begin{equation}
\label{Bos6}
H_2 = v\sum_{q>0} q (\alpha_{2q}^{R\dagger}\alpha_{2q}^R + \alpha_{2q}^{L\dagger}\alpha_{2q}^L ) 
\end{equation}
with a renormalized velocity $v=v_F\sqrt{(1+\bar g_{4,2})^2-\bar
  g_{2,2}^2}$. For the Loschmidt amplitude this leads to
\begin{equation}
\label{Bos7}
Z(z)=\prod_{q>0}\frac{1-\overline K^2}{1-\overline K^2\e^{-2zvq}}\;\mbox{with}\; \overline K=\frac{K_1-K_2}{K_1+K_2}  .
\end{equation}
We see that for a quench within the Luttinger model, $Z(z)$ does not
have any zeros in the complex plane.

For the XXZ model \eqref{model} with $h_\textrm{st}=0$, integrability
makes it possible to determine the Luttinger parameter $K$ and the
velocity $v$ exactly as a function of the anisotropy $\Delta$:
\begin{equation}
\label{Bos8}
v=\frac{\pi}{2}\frac{\sqrt{1 - \Delta^2}}{\arccos\Delta}\, , \quad K=\frac{\pi}{2(\pi - \arccos\Delta)}\, .
\end{equation}
We can thus try to compare the return rate\cite{Cazalilla,SachdevaNag,DoraPollmann} 
\begin{equation}
\label{Bos9}
l_\textrm{Bos}(t,\Lambda)=-\frac{1}{2\pi}\int_0^\Lambda\ln\left\{\frac{\left(1-\overline K^2\right)^2}{1+\overline K^4-2\overline K^2\cos(2tvq)}\right\} 
\end{equation}
with numerical data for quenches within the critical phase. However,
it is important to stress that sudden quenches are in general not
low-energy problems and we therefore cannot expect Luttinger liquid
theory to work. Only for small quenches, where a small amount of
energy is put into the system, might \eqref{Bos9} yield a reasonable
description. This problem is evident in Eq.~\eqref{Bos9} where we have
to use some cutoff scheme to make the integral convergent at short 
wavelengths. The result \eqref{Bos9} is thus nonuniversal and depends
on the properties of the microscopic model on the scale of the lattice 
constant. Here, we have simply introduced a hard cutoff $\Lambda$,
which we will use as a fitting parameter when comparing with the
numerical data in the next section.


\section{Numerical results}
\label{Numerics}
We now investigate the model \eqref{model} using the LCRG algorithm
described in Sec.~\ref{implementation}.
\subsection{Integrable case}
We will first concentrate on quenches where the Hamiltonian in the
time evolution operator $U(t)$ is given by the integrable XXZ model,
Eq.~\eqref{model} with $h_\textrm{st}=0$. For a chain of length $N$
this model has $N$-many local conserved charges, i.e., conserved
charges which can be written as a sum over a density operator with
finite support.\cite{GrabowskiMathieu,HubbardBook} In addition, it has
been recently shown that in the critical regime, $-1<\Delta< 1$, a set
of quasilocal conserved charges exist.\cite{Prosen,ProsenIllievski}
These local and quasilocal conservation laws do affect the
thermalization of a quantum system at long times after a quantum
quench and have to be included in the statistical
ensemble.\cite{RigolDunjko,FagottiEssler,SirkerKonstantinidis} In the
XXZ model, the thermal current is, for example, one of the local
conserved charges. If we prepare the system in an initial state
$|\Psi_0\rangle$ with a finite thermal current then this current will
remain nonzero for all times in the time evolved state
$|\Psi(t)\rangle$ while it would decay in a generic quantum system.
From this example it is obvious that the Loschmidt amplitude,
$\langle\Psi_0|\Psi(t)\rangle$, can behave very differently depending
on whether or not the Hamiltonian in the time evolution operator is
integrable. It therefore makes sense to consider the integrable and
nonintegrable cases separately.

\subsubsection{Quenches within the critical phase}
We start by considering small quenches in the Luttinger liquid phase
which can be compared with the prediction by bosonization. To obtain
the ground state, we use the projection method described in
Sec.~\ref{implementation}. The relative error, $\Delta
e_\textrm{rel}=|(e_0-e_\textrm{BA})/e_\textrm{BA}|$, in the
numerically obtained ground-state energy $e_0$ in comparison to the
exact one, $e_\textrm{BA}$, known from the Bethe ansatz\cite{tak99} is
given in Table \ref{Tab1}.

\begin{figure}
 \includegraphics*[width=1.0\columnwidth]{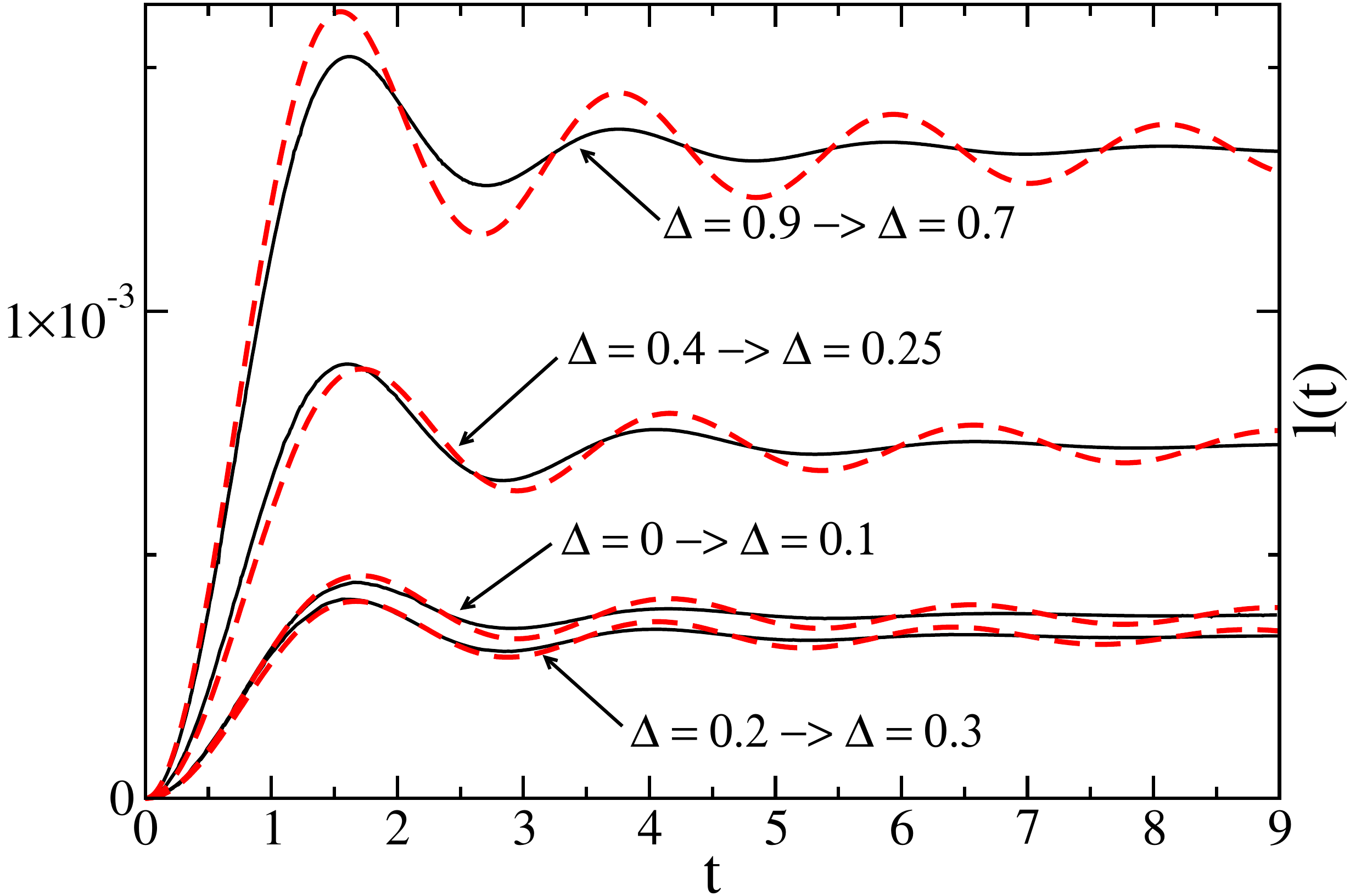}
 \caption{(Color online) Comparison between LCRG data (solid lines)
   and the bosonization result, $l_\textrm{Bos}(t,\Lambda)$ (dashed
   lines), with fit parameters $\Lambda$ and a velocity shift $v_0$ as
   given in table \ref{Tab1}.}
 \label{Fig_Bos1}
 \end{figure}

In the long-time limit bosonization predicts that the return rate
$l(t)=f(\im t)+f(-\im t)$ saturates at\cite{DoraPollmann}
 \begin{equation}
 \label{Bos8_2}
 \lim_{t\to\infty}l(t)=-\frac{\Lambda}{\pi}\ln\left(1-\overline K^2\right).  
 \end{equation}
 By demanding that this agrees with the numerically found saturation
 value, we ``fix'' the cutoff $\Lambda$. The values that we find
 from this condition are also given in Table \ref{Tab1}. In addition,
 we allow for a shift in the oscillation frequency by replacing $v\to
 v+v_0$ in Eq.~\eqref{Bos9} with $v_0$ used as an additional fitting
 parameter. Such a shift can be expected due to irrelevant operators
 that have not been taken into account in the Luttinger model.  The
 results of such fits are shown in Fig.~\ref{Fig_Bos1} and the fit
 parameter $v_0$ is given in Table \ref{Tab1}. Note that the Luttinger
 parameters $K_1$ and $K_2$ as well as the velocity $v$ of the
 collective excitations in the final Hamiltonian are fixed by
 Eq.~\eqref{Bos8}.
\begin{table}
\begin{ruledtabular}
\begin{tabular}{c|c|c|c|c}
$\Delta_\textrm{ini}$ & $\Delta e_\textrm{rel}$ & $\Delta_\textrm{fin}$ & $\Lambda$ & $v_0$ \\ \hline\hline

0 & $2.5\cdot 10^{-3}$ & 0.1 & 1.235 & 0.0 \\\hline
0.2 & $1.2\cdot 10^{-3}$ & 0.3 & 1.3 & -0.15 \\\hline
0.4 & $1.3\cdot 10^{-3}$ & 0.25 & 1.305 & -0.15 \\\hline
0.9 & $8.6\cdot 10^{-3}$ & 0.7 & 0.89 & 0.22 
\end{tabular}
\caption{\label{Tab1} The fit parameters $\Lambda$ and $v_0$ used in
  the comparison of the numerical data for the return rate $l(t)$ and
  the bosonization result \eqref{Bos9}, see Fig.~\ref{Fig_Bos1}.}
\end{ruledtabular}
\end{table}

While such fits are in reasonable agreement with the numerical data,
we want to stress that the predictive power of the bosonization result
is limited. In particular, the frequency of the oscillations in $l(t)$
depends on the long wavelength cutoff $\Lambda$. Furthermore, the
numerical data seem to show that the oscillations are much stronger
damped than predicted at this level by bosonization with the damping
becoming more pronounced the larger the interaction $\Delta$ is.

\subsubsection{Quenches across the first-order transition}
The XXZ model with $h=h_\textrm{st}=0$ has a first-order transition
between the Luttinger liquid and the ferromagnetic phase at
$\Delta=-1$. For a quench starting in the FM phase, we have $|Z(z)|=1$
because the FM state is an eigenstate of the XXZ Hamiltonian for all
interaction strengths $\Delta$. This is a trivial example showing that
a phase transition does not necessarily imply the existence of Fisher
zeros or nonanalyticities in the return rate.

\begin{figure}
  \includegraphics*[width=1.0\columnwidth]{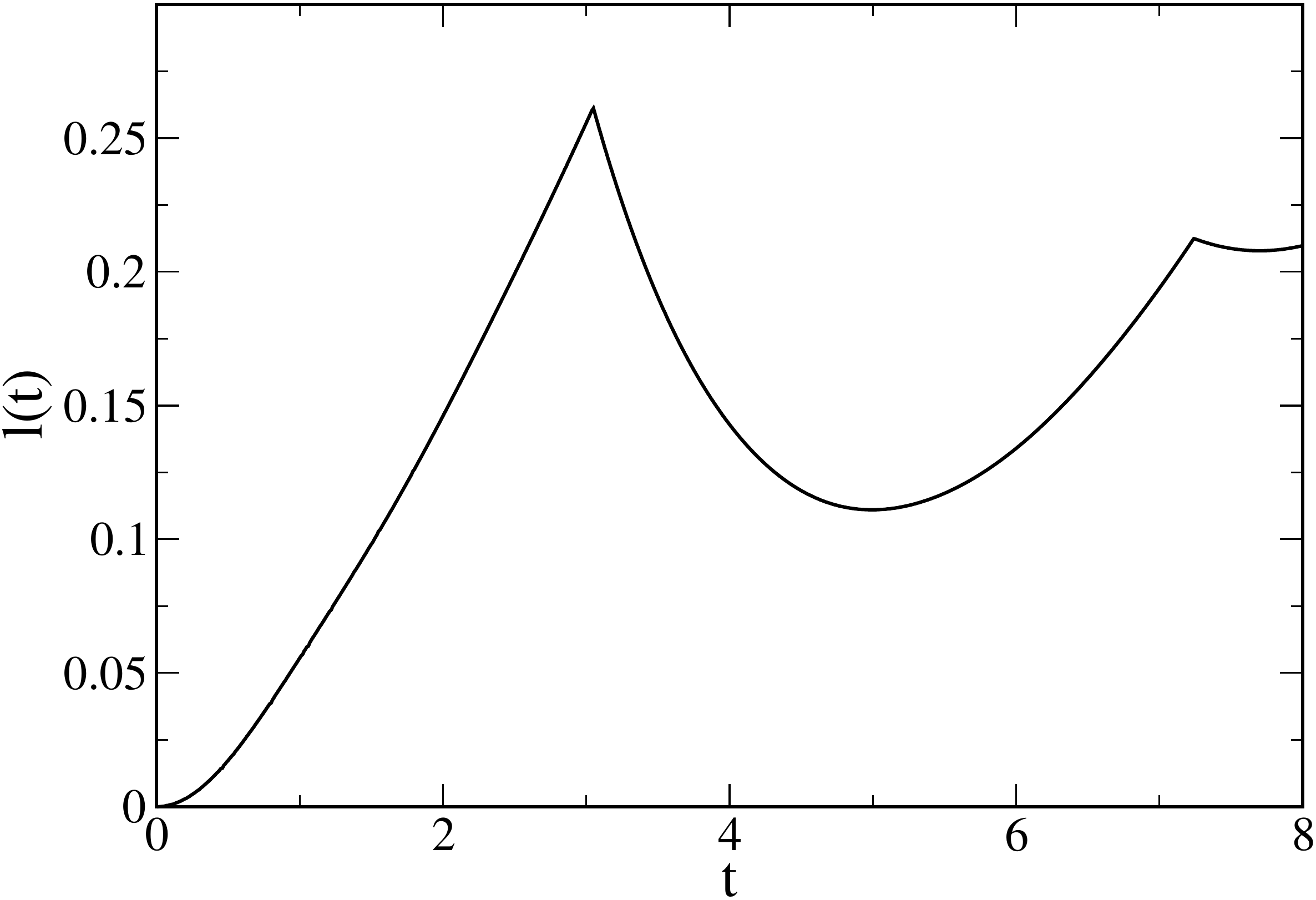}
 \caption{(Color online) Return rate $l(t)$ for a quench from the
   Luttinger liquid phase, $\Delta=-0.8$, into the ferromagnetic
   phase, $\Delta=-2.0$.}
 \label{Fig_Int_LL_FM}
 \end{figure}

Nontrivial results are, on the other hand, obtained for quenches from
the LL into the FM phase. Numerically, the calculations for such
quenches are again demanding because the projection method described
in Sec.~\ref{implementation} to obtain the ground state converges
rather slowly in a gapless phase. In Fig.~\ref{Fig_Int_LL_FM} we show
$l(t)$ for such a quench where we have obtained $\Delta
e_\textrm{rel}=4\times 10^{-4}$ for the ground-state energy of the
initial state.
 Two cusps in $l(t)$ are clearly visible in this case indicating a
 dynamical phase transition as defined by the divergence of
 $\overline\xi_k$, see Eq.~\eqref{corr_LE}.

\subsubsection{Quenches across the BKT transition}
\label{BKT_int}
\begin{figure}
 \includegraphics*[width=1.0\columnwidth]{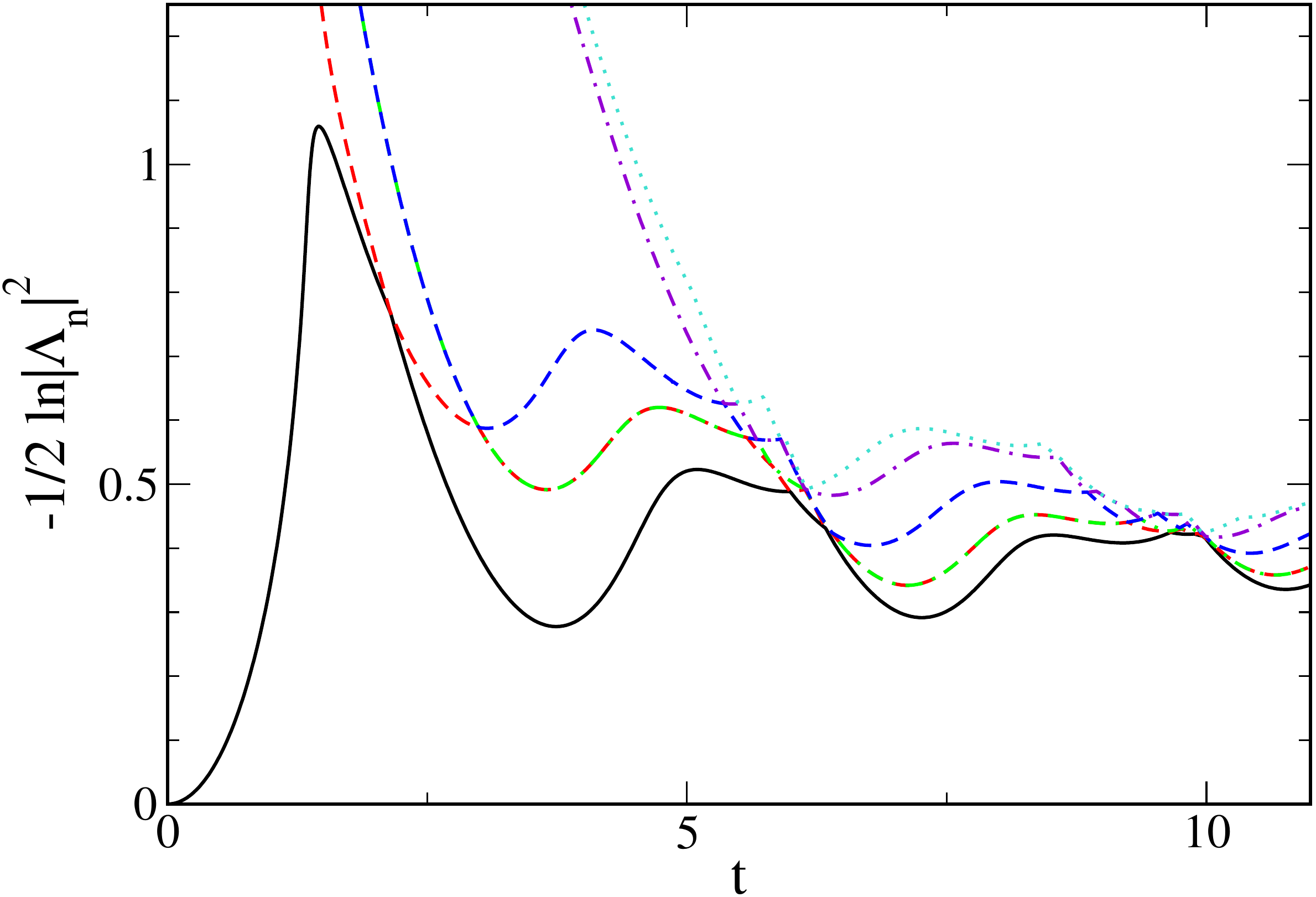}
 \caption{(Color online) Return rate $l(t)$ and next-leading
   eigenvalues, $-(\ln|\Lambda_n|^2)/2$, for a quench from the AFM to
   the LL phase. Here $\Delta_\textrm{ini}=6$, with the initial state
   being polarized, and $\Delta_\textrm{fin}=-0.6$.}
 \label{Fig_Int_AF_LL}
 \end{figure}

\begin{figure}
 \includegraphics*[width=1.0\columnwidth]{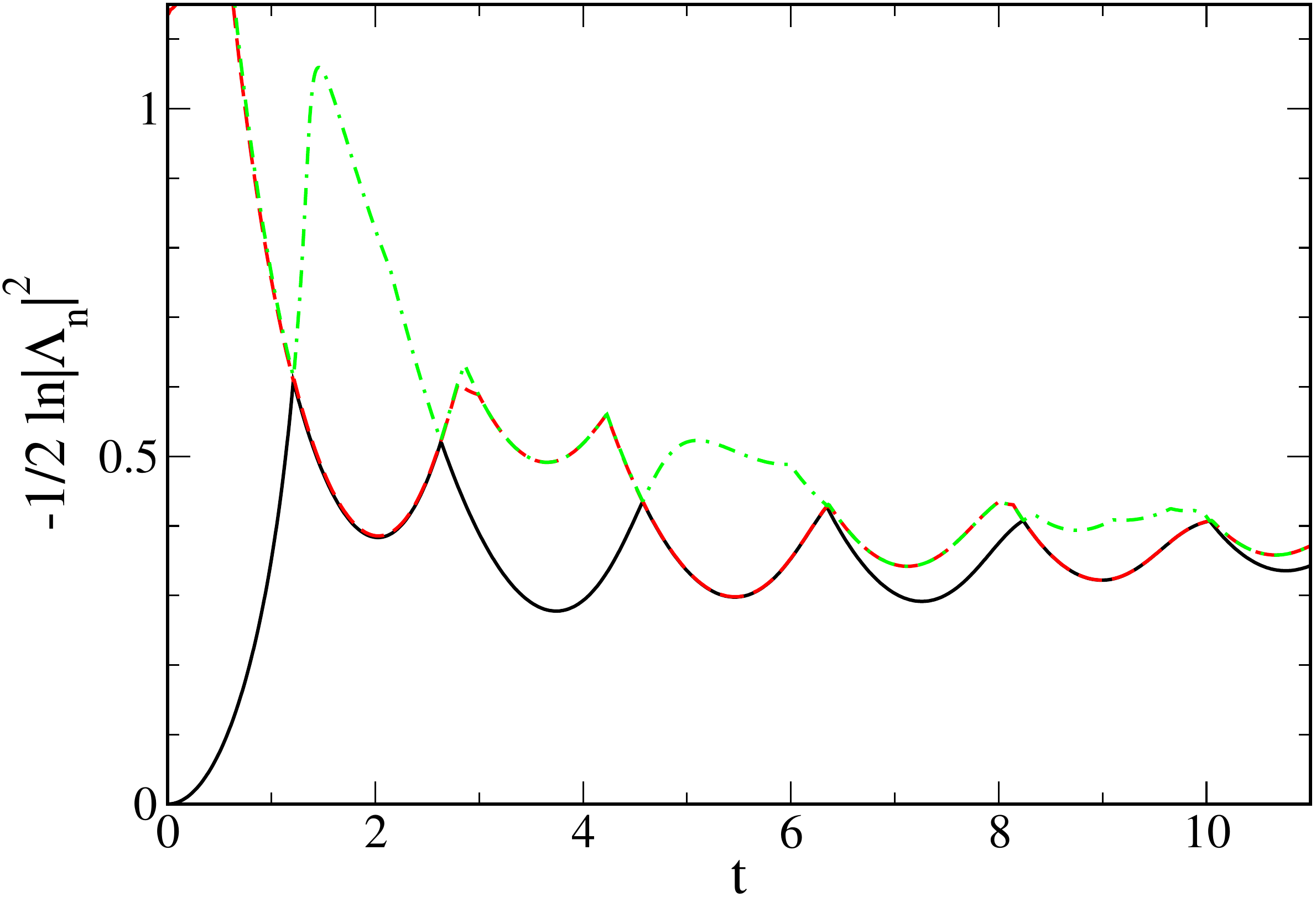}
 \caption{(Color online) Same quench as shown in
   Fig.~\ref{Fig_Int_AF_LL} but with a symmetric initial state. The
   eigenvalues shown are twofold degenerate.}
 \label{Fig_Int_AF_LL2}
 \end{figure}

At $\Delta=1$, the XXZ model in zero field shows a BKT transition
between the LL and the AFM phase driven by a marginal Umklapp
scattering term. The excitation gap on the AFM site is opening up
exponentially slowly, $\Delta E\sim\exp[-1/(\Delta-1)]$, and
derivatives of the energy do not show any singularities at finite
order. If cusps in the return rate are indeed related to quantum phase
transitions one might therefore wonder if they will also be present
for such an infinite order transition. In Figs.~\ref{Fig_Int_AF_LL} 
and~\ref{Fig_Int_AF_LL2}, we present an example for a quench from the
AFM into the LL phase. As an initial state, we can either take the
ground state where the $S^z_j\to -S^z_j$ symmetry is broken (polarized
state) or a state where this symmetry is preserved (symmetric state)
while the $U(1)$ symmetry (rotations in the $x$-$y$ plane) always
remains intact.
 In both cases, the leading and next-leading eigenvalues cross, leading
 to cusps in the return rate. The positions and the form of the cusps
 is, however, very different depending on whether we choose the
 polarized or the symmetric initial state. A similar dependence on
 the symmetry of the initial ground state has also been observed
 previously for generalized Ising models\cite{KarraschSchuricht} and
 in the quench to the free fermion point discussed in
 Sec.~\ref{FreeFermions}.

\begin{figure}
 \includegraphics*[width=1.0\columnwidth]{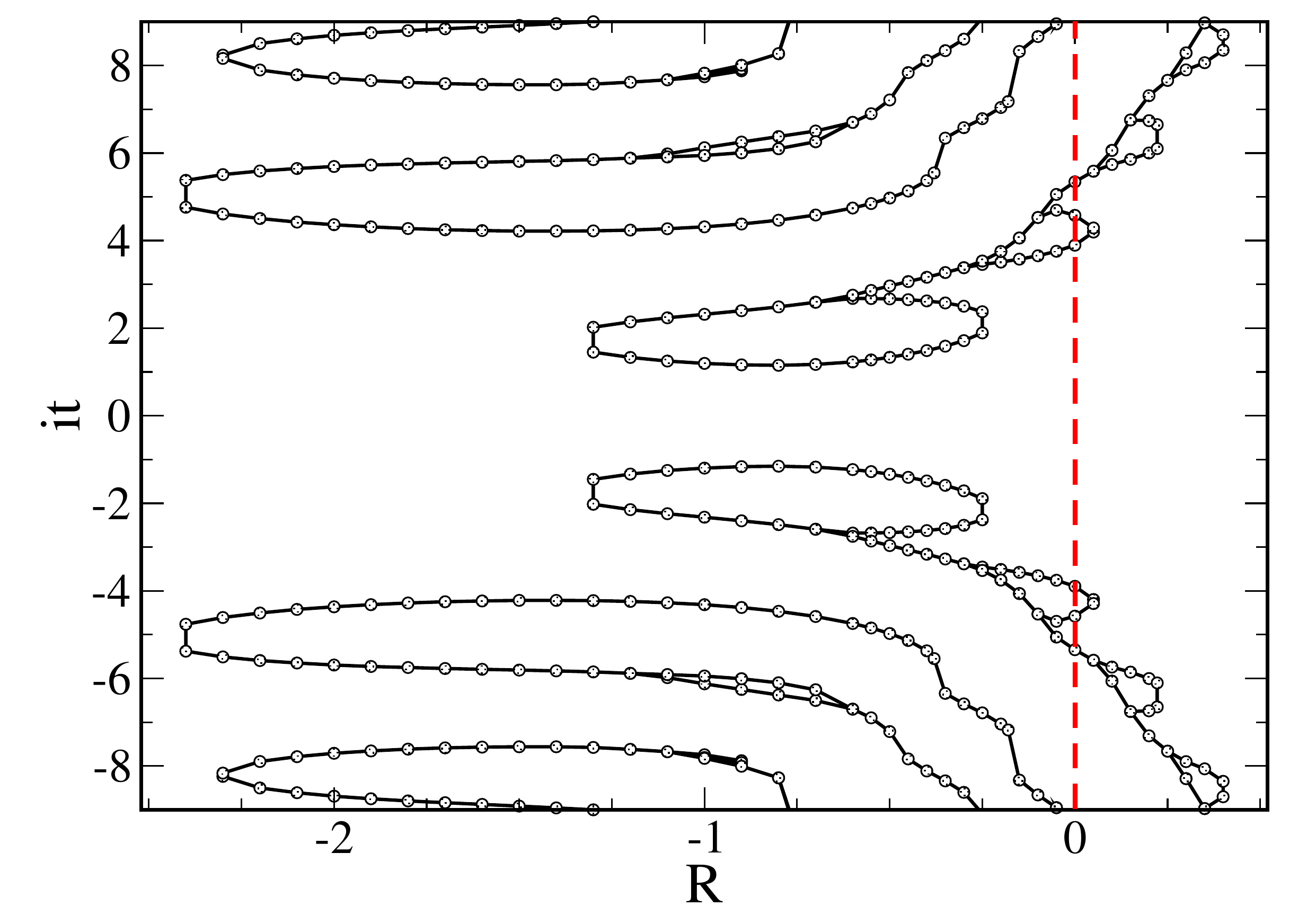}
 \caption{(Color online) Fisher zeros for a quench from the
   AFM into the LL phase with $\Delta_\textrm{ini}=6$ and
   $\Delta_\textrm{fin}=0.8$. The lines of Fisher zeros cut the $R=0$
   axis leading to nonanalyticities in the return rate $l(t)$. 
   Lines are a guide to the eye. }
 \label{Fig_Int_AF_LL_FZ}
 \end{figure}

 In addition to the return rate, we can also numerically obtain the
 Fisher zeros for a quench from the AFM into the LL phase. An example
 is shown in Fig.~\ref{Fig_Int_AF_LL_FZ} where the initial state is
 a symmetric state.
 The lines of Fisher zeros now look much more complicated than for the
 quench in the transverse Ising case shown in Fig.~\ref{Fig_Ising}.
 They form lines which seem to close for $R<0$. Furthermore, we find a
 number of ``bifurcation points'' where a line of zeros splits into two
 or where two lines join. In the considered case, the lines of Fisher
 zeros cross the $R=0$ axis at times $t_c$ where the return rate shows
 nonanalyticities.

\subsubsection{Quenches in the AFM phase}
\label{counterex1} 
\begin{figure}
 \includegraphics*[width=1.0\columnwidth]{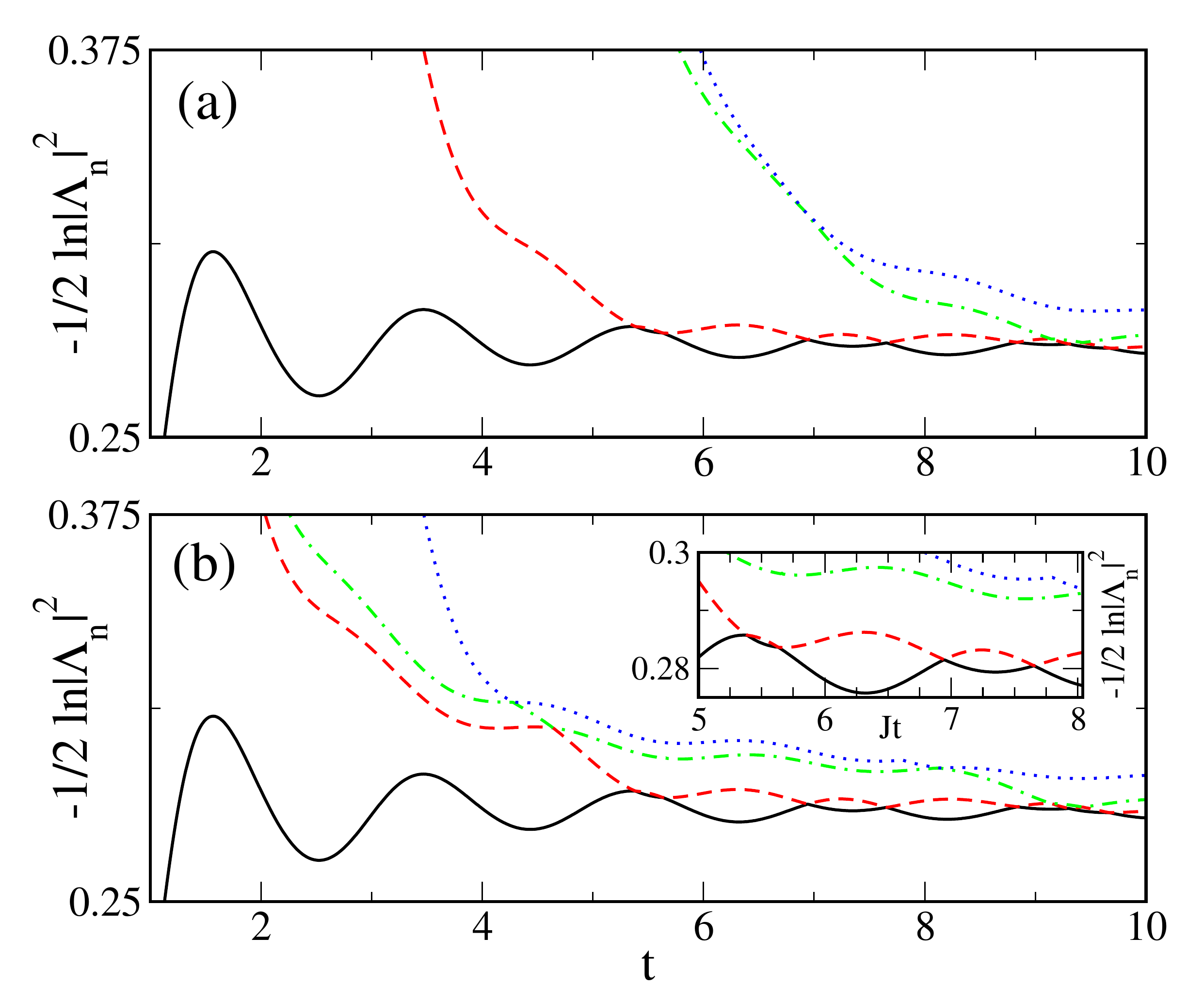}
 \caption{(Color online) (a) Quench from the polarized initial state
   with $\Delta_\textrm{ini}=20000$ and $\Delta_\textrm{fin}=1.2$
   which are both in the AFM phase.  Contrary to what might have been
   expected, $l(t)$ shows cusps. (b) Same as in (a) for the
   symmetric initial state.  All eigenvalues are twofold degenerate.
   The inset shows the regime where the first crossings of the leading
   and a next-leading eigenvalue occur.}
 \label{Fig_Int_AF_AF}
 \end{figure}

Next, we consider quenches within the AFM phase.  Because no phase
transition line is crossed in this case, one might expect that the
return rate is analytic. We will show, however, that this expectation
is wrong in general. Our first example is presented in
Fig.~\ref{Fig_Int_AF_AF} where we show the return rate as well as
next-leading eigenvalues of the transfer matrix for a quench where the
initial state is either polarized are symmetric.
 In both cases, we clearly observe nonanalyticities in $l(t)$. While
 the return rate itself barely changes, the spectrum of the transfer
 matrix is different for the two initial states. The crossings between
 the leading and next-leading eigenvalue are clearly resolved
 numerically, in particular at later times, making this an obvious
 counterexample in the integrable case with regard to the hypothesis
 that cusps in $l(t)$ are always related to equilibrium quantum
 critical points.

\begin{figure}
 \includegraphics*[width=1.0\columnwidth]{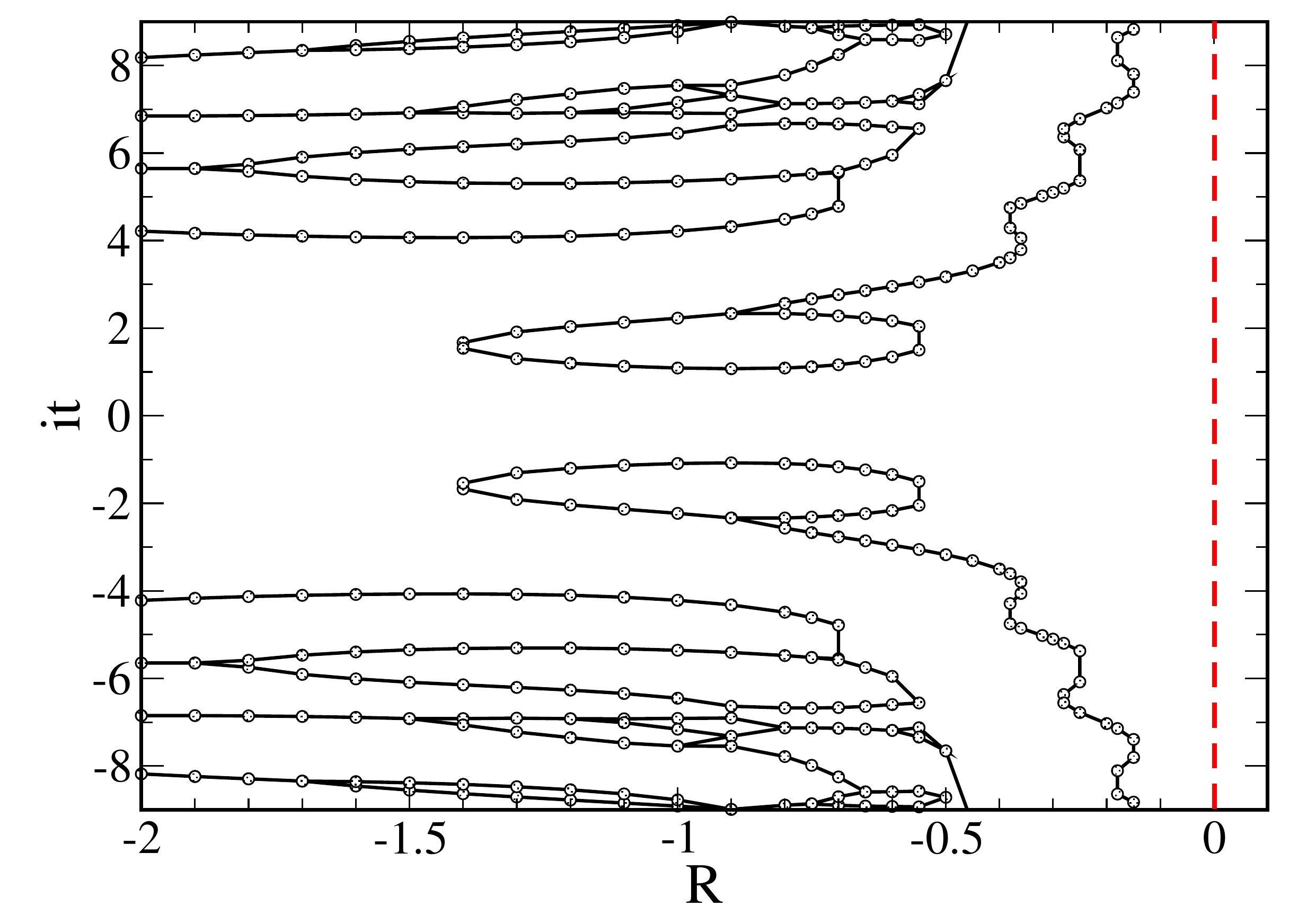}
 \caption{(Color online) Fisher zeros for a quench within the
   AFM phase with $\Delta_\textrm{ini}=6$ (symmetric) and
   $\Delta_\textrm{fin}=1.1$. The Fisher zeros might cut the
   $R=0$ axis at longer times. Lines are a guide to the eye.}
 \label{Fig_Int_AF_FZ}
 \end{figure}

 Second, we show in Fig.~\ref{Fig_Int_AF_FZ} the lines of Fisher zeros
 for a different quench in the AFM phase. 
 The structure of the Fisher zeros is again fairly complicated. Of
 particular interest is the line of Fisher zeros which approaches the
 $R=0$ axis at the longest simulation times. Here we finally cannot
 resolve whether or not this line will cut the $R=0$ axis at long
 times.  What we do observe, however, is that the time where $l(t)$
 shows the first cusp increases with decreasing $\Delta_\textrm{ini}$
 while keeping $\Delta_\textrm{fin}$ fixed. This suggests that
 nonanalyticities are a {\it generic feature} for quenches within the
 AFM phase and that it is merely the limited simulation time which
 allows us to resolve cusps for the quench shown in
 Fig.~\ref{Fig_Int_AF_AF}, while we cannot resolve them for the case
 shown in Fig.~\ref{Fig_Int_AF_FZ}.

\subsubsection{Quenches across the BKT transition driven by a staggered field}
\begin{figure}
 \includegraphics*[width=1.0\columnwidth]{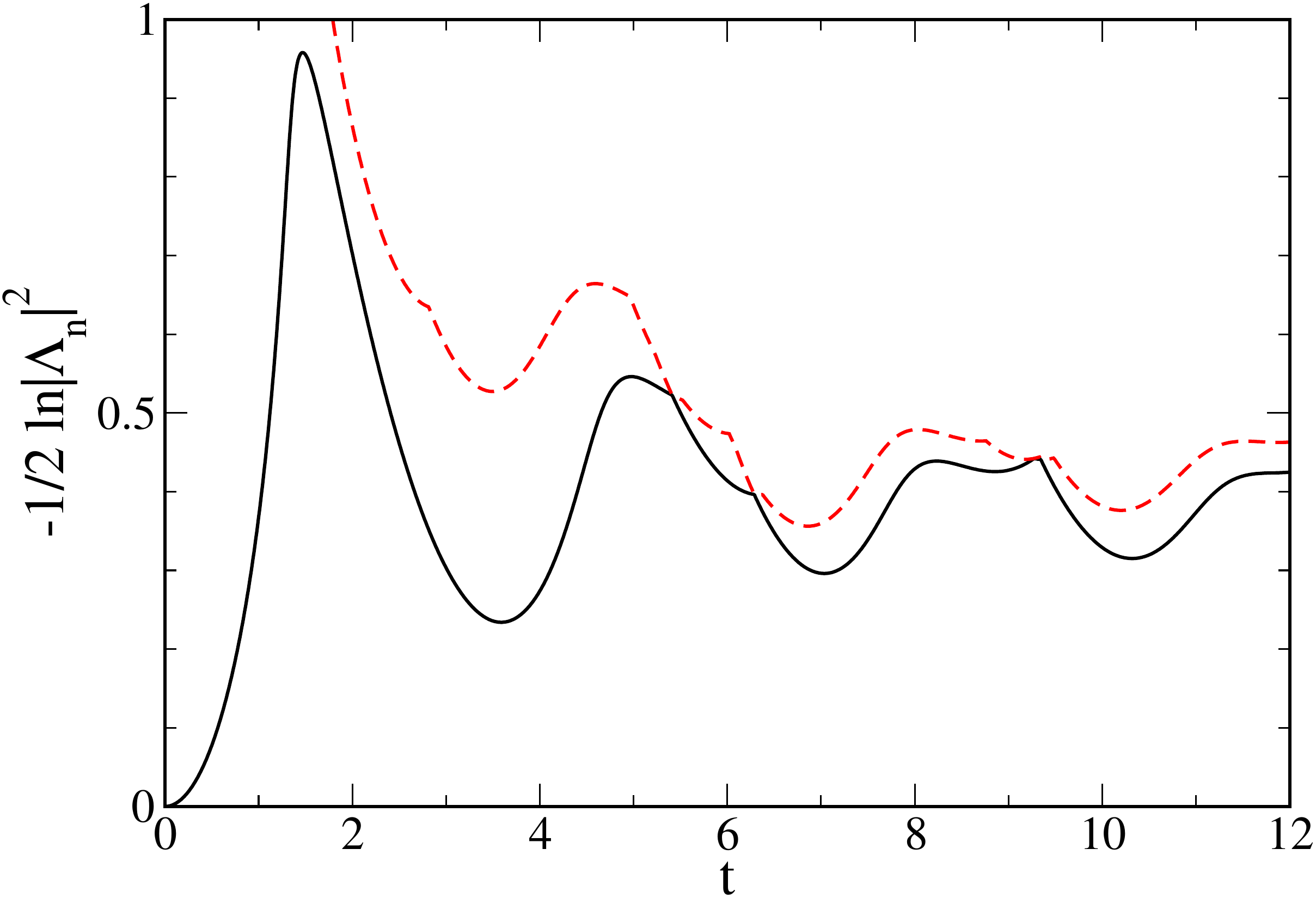}
 \caption{(Color online) Return rate $l(t)$ and next-leading
   eigenvalues, $-(\ln|\Lambda_n|^2)/2$, for a quench across the BKT
   transition from the AFM to the LL phase. Here
   $\Delta_\textrm{ini}=\Delta_\textrm{fin}=-0.8$,
   $h_\textrm{st,ini}=1.0$, and
   $h_\textrm{st,fin}=0.0$.}
 \label{Fig_Int_AF_LL_staggfield}
 \end{figure}

Finally, we want to consider quenches where the initial state is the
ground state of the nonintegrable Hamiltonian \eqref{model} with
$h_\textrm{st}\neq 0$, while the time evolution is still integrable.
For the quench shown in Fig.~\ref{Fig_Int_AF_LL_staggfield}, we have
chosen the parameters such that the critical line between the AFM and
the LL phase is crossed, see the phase diagram in
Fig.~\ref{Fig_phase_diag}(b).
 As might be expected, the result looks qualitatively similar to the
 quench from the AFM to the LL phase shown in Fig.~\ref{Fig_Int_AF_LL}
 without the staggered field but with the initial state being
 polarized.  

\subsection{Nonintegrable case}
Now we turn to quenches where the time-evolving Hamiltonian is given
by \eqref{model} with $h_\textrm{st}\neq 0$. In this case the model is
nonintegrable and the total magnetization and the Hamiltonian itself
are the only local conserved charges. Subsystems of an infinite XXZ
chain with staggered field are therefore expected to thermalize at
long times after the quench.
\subsubsection{Quenches across the second-order transition driven by the staggered field}
\begin{figure}
 \includegraphics*[width=1.0\columnwidth]{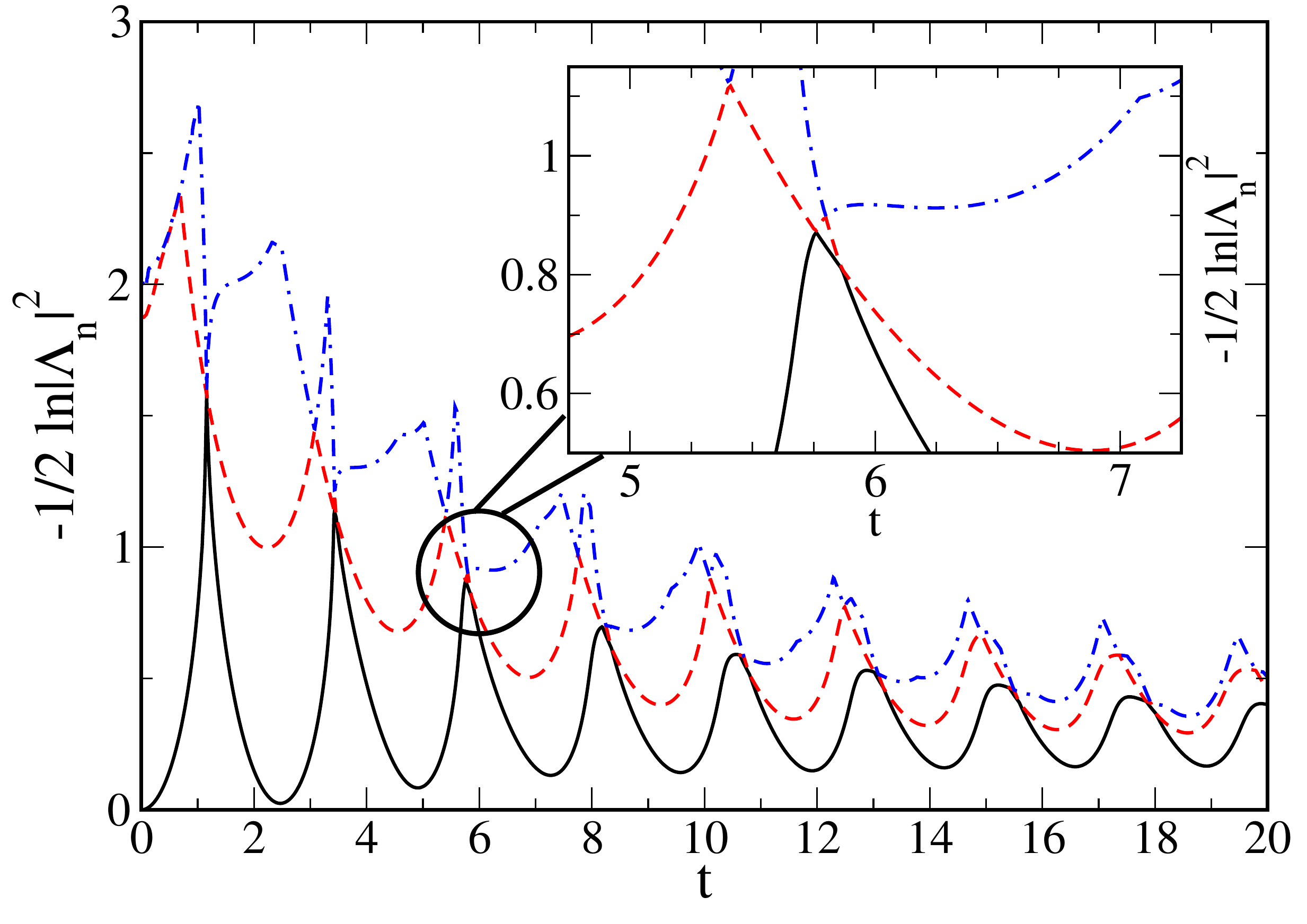}
 \caption{(Color online) Return rate $l(t)$ (solid line) and
   next-leading eigenvalues, $-(\ln|\Lambda_n|^2)/2$, for a quench
   across the second order transition between the AFM phases with
   $\Delta_\textrm{ini}=\Delta_\textrm{fin}=0.1$, 
   $h_\textrm{st,ini}=0.5$ and $h_\textrm{st,fin}=-0.5$.}
 \label{Fig_NonInt_AF_AF}
 \end{figure}

For $\Delta > -1/\sqrt{2}$, there is a second-order phase transition
when changing the sign of the staggered field, see
Fig.~\ref{Fig_phase_diag}(b). The leading eigenvalues for a quench
across this transition are shown exemplarily in
Fig.~\ref{Fig_NonInt_AF_AF} for a weakly interacting system. 
 Nonanalyticities in $l(t)$ are present, however, the regions where
 crossings between the leading and the next-leading eigenvalue occur
 are very narrow in time so that it might be possible to find
 parameters where the second-order transition is crossed in the quench
 but $l(t)$ stays analytic. 

\begin{figure}
 \includegraphics*[width=1.0\columnwidth]{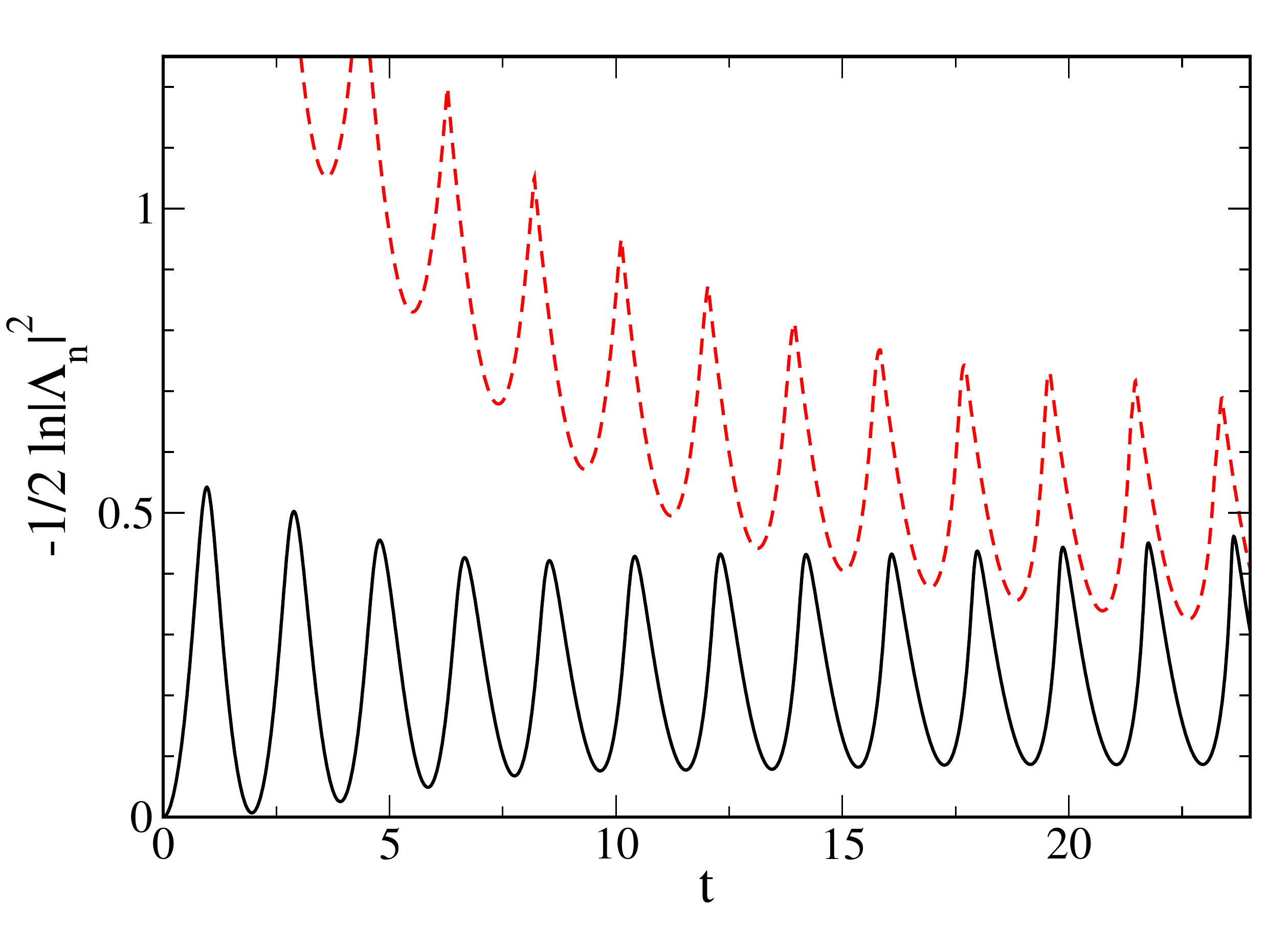}
 \caption{(Color online) Return rate $l(t)$ (solid line) and
   next-leading eigenvalues, $-(\ln|\Lambda_n|^2)/2$, for a quench
   across the second order transition between the AFM phases with
   $\Delta_\textrm{ini}=\Delta_\textrm{fin}=0.1$,
   $h_\textrm{st,ini}=0.7$ and $h_\textrm{st,fin}=-0.7$.}
 \label{Fig_NonInt_AF_AF2}
 \end{figure}

 By quenching the field instead from $h_\textrm{st,ini}=0.7$ to
 $h_\textrm{st,fin}=-0.7$, we indeed no longer find any cusps in the
 numerically accessible time range, see Fig.~\ref{Fig_NonInt_AF_AF2}.
 Obviously, we cannot rule out that nonanalyticities will show up at
 longer times. What we can say, however, is that apart from the
 counterexamples where a quantum phase transition is crossed but
 $l(t)=0$ discussed before---i.e., quenches from the FM phase or
 quenches where only $h$ is changed---we also find quenches across
 equilibrium transitions leading to nontrivial dynamics but with
 $l(t)$ being analytic at least for very long times.

\subsubsection{Quenches across the BKT transition driven by a staggered field}
\begin{figure}
 \includegraphics*[width=1.0\columnwidth]{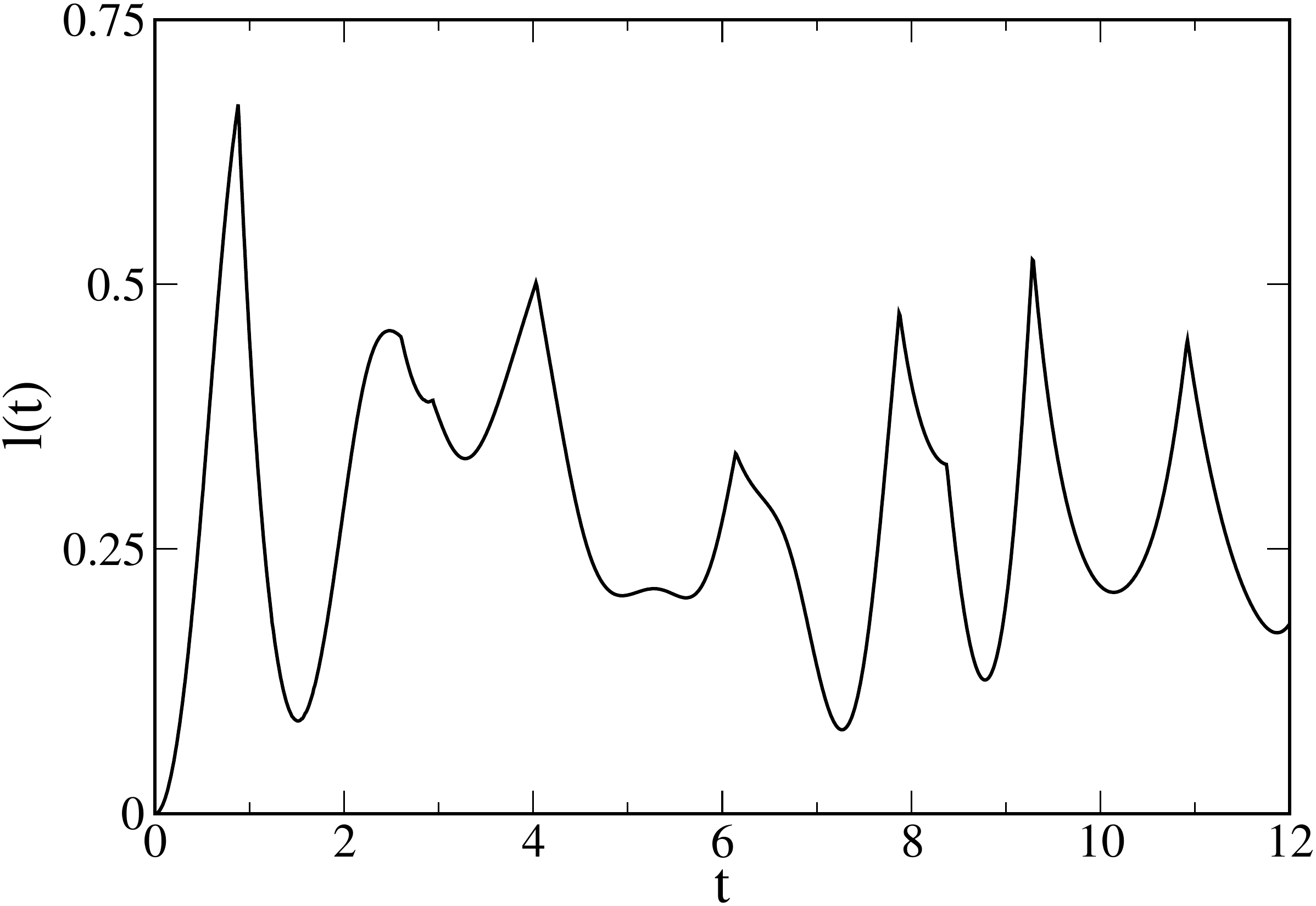}
 \caption{(Color online) Return rate $l(t)$ for a quench across the
   BKT transition from the LL to the AFM phase. Here,
   $\Delta_\textrm{ini}=\Delta_\textrm{fin}=-0.8$,
   $h_\textrm{st}^\textrm{ini}=0.0$, and
   $h_\textrm{st}^\textrm{fin}=1.0$.}
 \label{Fig_NonInt_LL_AF_staggfield}
 \end{figure}

It is also possible to cross a BKT transition line by changing the
staggered field. In Fig.~\ref{Fig_NonInt_LL_AF_staggfield}, the return
rate for such a quench is shown where we start with the same Luttinger
liquid ground state as in Fig.~\ref{Fig_Int_LL_FM} and quench into the
AFM phase.
 Cusps in $l(t)$ are clearly visible. Here, the numerically accessible
 time range is again quite limited because of the slowly converging
 projection onto the gapless ground state.

\subsubsection{Quenches within the AFM phase}
For the integrable model we discussed in Sec.~\ref{counterex1}
examples for quenches which do not cross an equilibrium phase
transition but still show nonanalyticities in the return rate. Here
we want to provide numerical evidence that such counterexamples also
exist in the nonintegrable model. Our starting point are the two
quenches in the integrable model from the AFM phase into the Luttinger
liquid phase across the BKT point at $\Delta=1$ shown in
Figs.~\ref{Fig_Int_AF_LL} and \ref{Fig_Int_AF_LL2}. If we keep the
same values for $\Delta_\textrm{ini}$ and $\Delta_\textrm{fin}$ and
add a very small staggered field, we might expect that the return rate
$l(t)$ will look very similar to the quench in
Fig.~\ref{Fig_Int_AF_LL} where we have used a polarized initial state.
The quench puts a large amount of energy into the system and small
changes to the low-energy spectrum will hardly matter. On the other
hand, any arbitrarily small staggered field introduces a cosine term
into the low-energy effective theory \eqref{Bos1} which is relevant
for $\Delta>-1/\sqrt{2}$. Quenches where both $\Delta_\textrm{ini}$
and $\Delta_\textrm{fin}$ are larger than $-1/\sqrt{2}$ therefore do
not cross an equilibrium phase transition anymore. If
nonanalyticities in $l(t)$ are always related to phase transitions,
this would suggest that they do not occur for these kind of quenches.

\begin{figure}
 \includegraphics*[width=1.0\columnwidth]{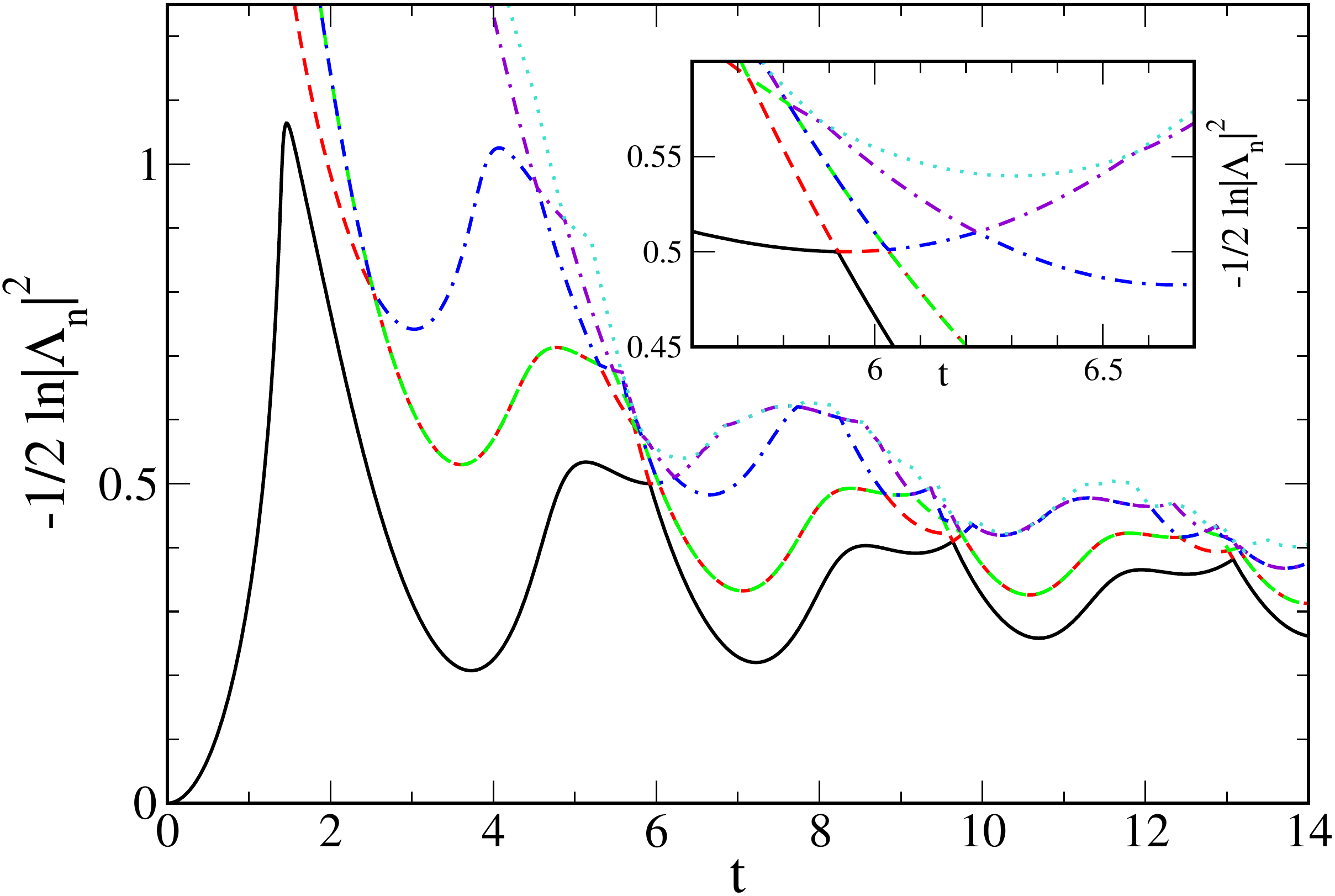}
 \caption{(Color online) Return rate $l(t)$ and next-leading
   eigenvalues, $-(\ln|\Lambda_n|^2)/2$, for a quench within the AFM
   phase. Here $\Delta_\textrm{ini}=6$, $\Delta_\textrm{fin}=-0.6$
   while the staggered field is kept constant,
   $h_\textrm{st}^\textrm{ini}=h_\textrm{st}^\textrm{fin}=0.05$. Note
   that $l(t)$ shows nonanalyticities although no phase transition is
   crossed. The inset shows the regime where the first crossing
   occurs.}
 \label{Fig_NonInt_AF_noPT}
 \end{figure}

\begin{figure}
 \includegraphics*[width=1.0\columnwidth]{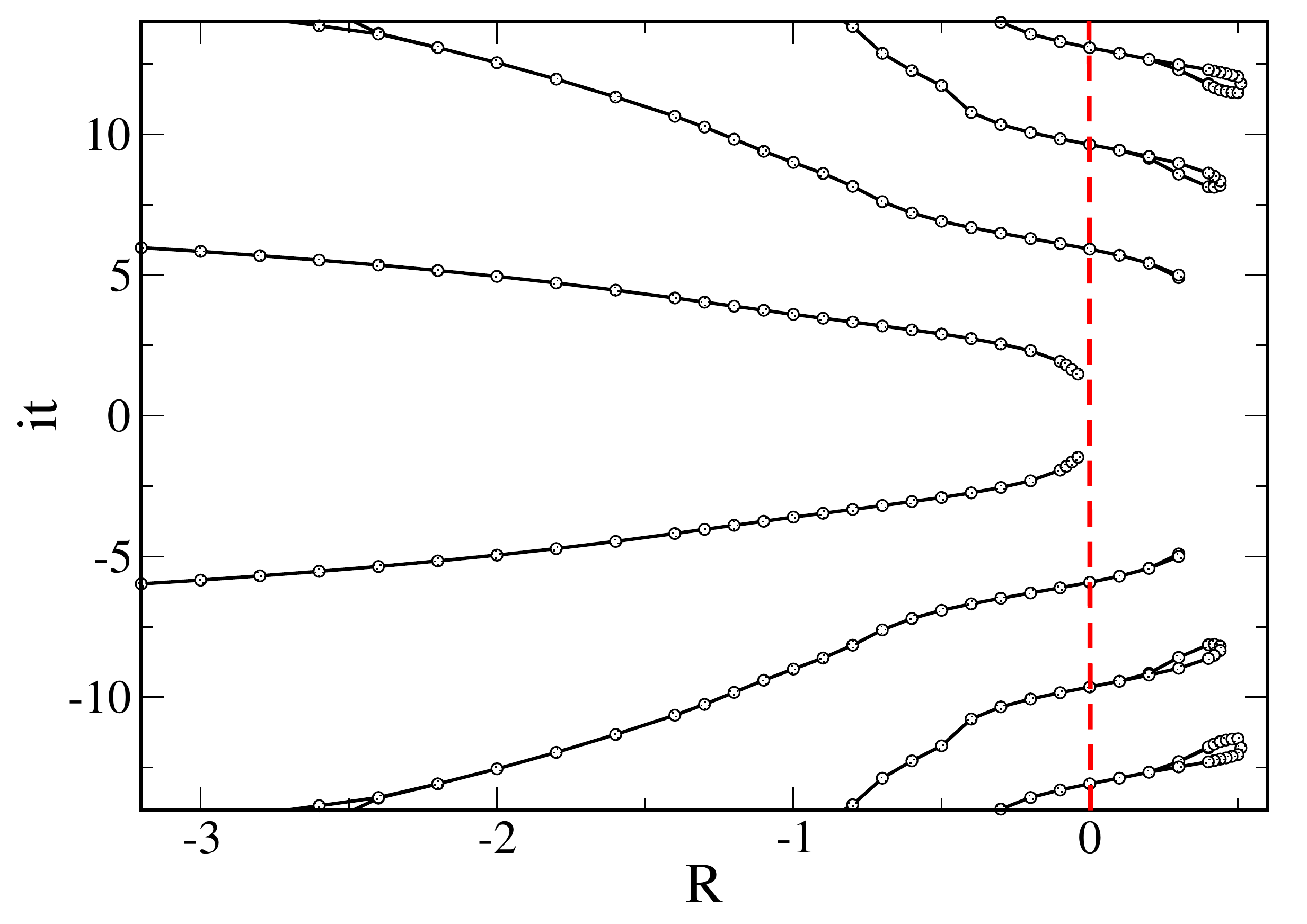}
 \caption{(Color online) Fisher zeros for the same quench as
   in Fig.~\ref{Fig_NonInt_AF_noPT}. The Fisher zeros cut the
   $R=0$ axis leading to the nonanalyticities in $l(t)$ shown in
   Fig.~\ref{Fig_NonInt_AF_noPT}. Lines are a guide to the eye.}
 \label{Fig_NonInt_AF_noPT_FZ}
 \end{figure}

From the numerical data in Fig.~\ref{Fig_NonInt_AF_noPT} it is obvious
that the first reasoning is correct. Adding a small staggered field
indeed hardly changes the return rate and Figs.~\ref{Fig_Int_AF_LL} and
\ref{Fig_NonInt_AF_noPT} do indeed look very similar.
 In addition, we can calculate the lines of Fisher zeros, which are
 shown in Fig.~\ref{Fig_NonInt_AF_noPT_FZ}.
 The line closest to the $t=0$ axis (the diagram is $t\to-t$
 symmetric) closely approaches the $R=0$ axis but does not cross. At
 the time $t\approx 1.36$ where it does get closest to the $R=0$ axis,
 the return rate, shown in Fig.~\ref{Fig_NonInt_AF_noPT}, has a
 sharp peak but remains analytic. The lines of Fisher zeros at larger
 times then do cross and the times where this happens are consistent
 with the nonanalyticities observed in $l(t)$ in
 Fig.~\ref{Fig_NonInt_AF_noPT}. The considered case is therefore an
 example for a quench within the same phase of a generic quantum model
 where, nevertheless, the Fisher zeros do cross the $R=0$ axis.

\section{Conclusions}
\label{Concl}
The generalized Loschmidt amplitude $Z(z)=\langle \Psi_0 |
\Psi(z)\rangle$ describes the overlap between an initial state
$|\Psi_0\rangle$ and the evolved state $|\Psi(z)\rangle
=\e^{-zH}|\Psi_0\rangle$ where $z$ is a complex number. For $z=\im t$,
this corresponds to real-time evolution, which is the experimentally
accessible case. Similar to studies of the canonical partition
function $Z_\textrm{th}(z)=\Tr\e^{-zH}$ it is, however, interesting to
study $Z(z)$ in the whole complex plane in order to understand the
analytic structure. The so-called Fisher zeros of the partition
function in the complex plane are important, because points where they
approach the real axis in the thermodynamic limit determine the
temperature where the corresponding thermodynamic potential becomes
nonanalytic, indicating a phase transition. It therefore seems to be
an excellent idea to carry this approach over to the Loschmidt echo
and ask if dynamical phase transitions can be defined
accordingly.\cite{HeylPolkovnikov}

For a $d$-dimensional quantum system, we can represent both
$Z_\textrm{th}$ and $Z(z)$ as a lattice path integral of a
$d+1$-dimensional classical model. The only difference is that the
boundary conditions along the $z$ direction are periodic for
$Z_\textrm{th}$ while they are fixed by the initial state for $Z(z)$.
For the classical system a transfer matrix can be defined in both
cases that evolves along the spatial direction. The free energy
$f(z)$ or the return rate, respectively, are then in the thermodynamic
limit fully determined by the eigenvalue $\Lambda_0$ of the transfer
matrix with largest modulus. Since eigenvalues can cross,
nonanalyticities in these quantities are possible. In both cases the
Fisher zeros are simply determined by the condition
$|\Lambda_0|=|\Lambda_n|$ where $\Lambda_n$ is the next-leading
eigenvalue with largest modulus, which is not degenerate with
$\Lambda_0$ except at the crossing point.

We have described in this work a numerical algorithm to calculate the
spectrum of the transfer matrix for $Z(z)$ in the one-dimensional
case. By comparing with exact results for the transverse Ising model,
we have shown that this algorithm can be used to precisely determine
the Fisher zeros in the complex plane. We have then concentrated on
the spin-$1/2$ XXZ model with a staggered magnetic field which we
used to switch between integrable and nonintegrable dynamics. For a
quench from a N\'eel state to the XX point (free fermions) we have
analytically shown that the Fisher zeros sensitively depend on the
initial state. While for a polarized N\'eel state as initial state,
$|\Psi_0\rangle
=|\uparrow\downarrow\uparrow\downarrow\,\cdots\rangle$, the Fisher
zeros are just single points that lie on the $z=\im t$ axis, the
Fisher zeros become closed, cigar-shaped lines in the complex plane
if we use instead the symmetric state $|\Psi_0\rangle
=(|\uparrow\downarrow\uparrow\downarrow\,\cdots\rangle +
|\downarrow\uparrow\downarrow\uparrow\,\cdots\rangle)/\sqrt{2}$. In
addition, we used bosonization for quenches within the Luttinger
liquid phase. The obtained result for the return rate $l(t)$, however,
has to be considered with care. A sudden quench generally leads to
particle-hole type excitations whose energies are only limited by the
bandwidth of the microscopic model so that the dynamics is no longer
dominated by low-energy excitations near the Fermi points. Only for
very small quenches might the bosonization approach be justified.
However, the bosonization result for $l(t)$ does even in this case
depend on the ultraviolet cutoff which we had to use as a fitting
parameter in our comparison with the numerical results. Allowing also
for a small frequency shift, we did find reasonable agreement with the
numerical data. 

In the rest of the paper, we discussed numerical data for the Fisher
zeros and return rates for various quenches in the XXZ with and
without staggered field in cases where analytical results are not
available. The main question we were trying to address is if
nonanalyticities in the return rate following a sudden quench are
always related to an equilibrium phase transition. For the XXZ model
in the integrable and nonintegrable case, we pointed out two trivial
counterexamples. (1) For any quench starting in the ferromagnetic
phase, we have $l(t)=0$ because the ferromagnetic state is an
eigenstate of the XXZ Hamiltonian for all parameters. (2) Any quench
where only the uniform field $h$ is changed leads to $l(t)=0$ because
the total magnetization always commutes with the Hamiltonian.  In both
cases, a crossing of an equilibrium phase transition in the quench does
not leave any trace in $l(t)$. We also found indications that quenches
in the staggered field across the second order transition between AFM
phases might not lead to cusps in $l(t)$ for certain parameters or at
least not in the numerically accessible time range. 

Even more important, we also found examples where no phase transition
is crossed but $l(t)$ nevertheless does show nonanalytic behavior. In
the integrable XXZ model, such counterexamples are found for quenches
within the antiferromagnetic phase from large values of anisotropy
$\Delta$ to $\Delta\gtrsim 1$. In the nonintegrable model with
staggered field, we did find such counterexamples again for quenches
within the antiferromagnetic phase where the staggered field is kept
constant and both the initial and the final values of anisotropy are
larger than $-1/\sqrt{2}$.

\acknowledgments The authors acknowledge support by the Collaborative
Research Centre SFB/TR49 and the Graduate School of Excellence MAINZ.
We are grateful to the Regional Computing Center at the University of
Kaiserslautern and the AHRP for providing computational resources and
support.


\end{document}